\documentclass[acmsmall]{acmart}

\usepackage{algorithmic}
\usepackage{graphicx}
\usepackage{textcomp}
\usepackage{xcolor}

\usepackage{hyperref}
\usepackage{booktabs}
\usepackage{subcaption}
\usepackage[algoruled,linesnumbered]{algorithm2e}
\makeatletter
\renewcommand{\@algocf@capt@plain}{above}% formerly {bottom}
\makeatother
\usepackage{bm}
\usepackage{multirow}
\usepackage{enumitem}
\usepackage{makecell}
\usepackage{rotating}
\newcommand{\model}{{RESUS}~}
\newcommand{\modelns}{{RESUS}}
\newcommand{\eat}[1]{}

\AtBeginDocument{%
  \providecommand\BibTeX{{%
    \normalfont B\kern-0.5em{\scshape i\kern-0.25em b}\kern-0.8em\TeX}}}

\setcopyright{acmcopyright}
\copyrightyear{2022}
\acmYear{2022}
\acmDOI{10.1145/3564283}

\acmJournal{JACM}
\acmVolume{37}
\acmNumber{4}
\acmArticle{1}
\acmMonth{9}

\begin{document}                        

\title{RESUS: Warm-Up Cold Users via Meta-Learning Residual User Preferences in CTR Prediction}


\author{Yanyan Shen}
\authornote{Yanyan Shen is the corresponding author.}
\affiliation{%
  \institution{Department of Computer Science and Engineering, Shanghai Jiao Tong University}
  % \streetaddress{}
  % \city{}
  \country{China}}
\email{shenyy@sjtu.edu.cn}

\author{Lifan Zhao}
\affiliation{%
	\institution{Department of Computer Science and Engineering, Shanghai Jiao Tong University}
	\country{China}}
\email{mogician233@sjtu.edu.cn}

\author{Weiyu Cheng}
\affiliation{%
  \institution{Department of Computer Science and Engineering, Shanghai Jiao Tong University}
  \country{China}}
\email{weiyu\_cheng@sjtu.edu.cn}

\author{Zibin Zhang}
\affiliation{%
  \institution{WeChat, Tencent}
  \country{China}}
\email{bingozhang@tencent.com}

\author{Wenwen Zhou}
\affiliation{%
  \institution{WeChat, Tencent}
  \country{China}}
\email{wendizhou@tencent.com}

\author{Kangyi Lin}
\affiliation{%
  \institution{WeChat, Tencent}
  \country{China}}
\email{plancklin@tencent.com}

\renewcommand{\shortauthors}{Shen, et al.}

\begin{abstract}
%Various approaches have been developed for Click-Through Rate (CTR) prediction which is an essential task in recommender systems. They aim to learn user preferences from historical behaviors to predict the probabilities of new clicks. 
%However, real-world recommender systems typically involve large numbers of cold users who have committed very few interaction records, and promoting CTR prediction performance on cold users remains a great challenge. 
Click-Through Rate (CTR) prediction on cold users is a challenging task in recommender systems. Recent researches have resorted to meta-learning to tackle the cold-user challenge, which either perform few-shot user representation learning or adopt optimization-based meta-learning. However, existing methods suffer from information loss or inefficient optimization process, and they fail to explicitly model global user preference knowledge which is crucial to complement the sparse and insufficient preference information of cold users. In this paper, we propose a novel and efficient approach named RESUS, which decouples the learning of global preference knowledge contributed by collective users from the learning of residual preferences for individual users. Specifically, we employ a shared predictor to infer basis user preferences, which acquires global preference knowledge from the interactions of different users. Meanwhile, we develop two efficient algorithms based on the nearest neighbor and ridge regression predictors, which infer residual user preferences via learning quickly from a few user-specific interactions. Extensive experiments on three public datasets demonstrate that our RESUS approach is efficient and effective in improving CTR prediction accuracy on cold users, compared with various state-of-the-art methods.
\end{abstract}

  


\maketitle

\section{Introduction}
%problem:
%Ctr prediction, conventional approaches, cold users. 
%very few clicking records make the ctr prediction of cold users suffer from low prediction accuracy.
%how to infer the preferences of these users? a prime example of this scenario is Wechat Article recommendation service.  

\begin{figure}[t]
	%	\captionsetup{labelfont=bf}
	\centering  
	\includegraphics[width=0.7\linewidth]{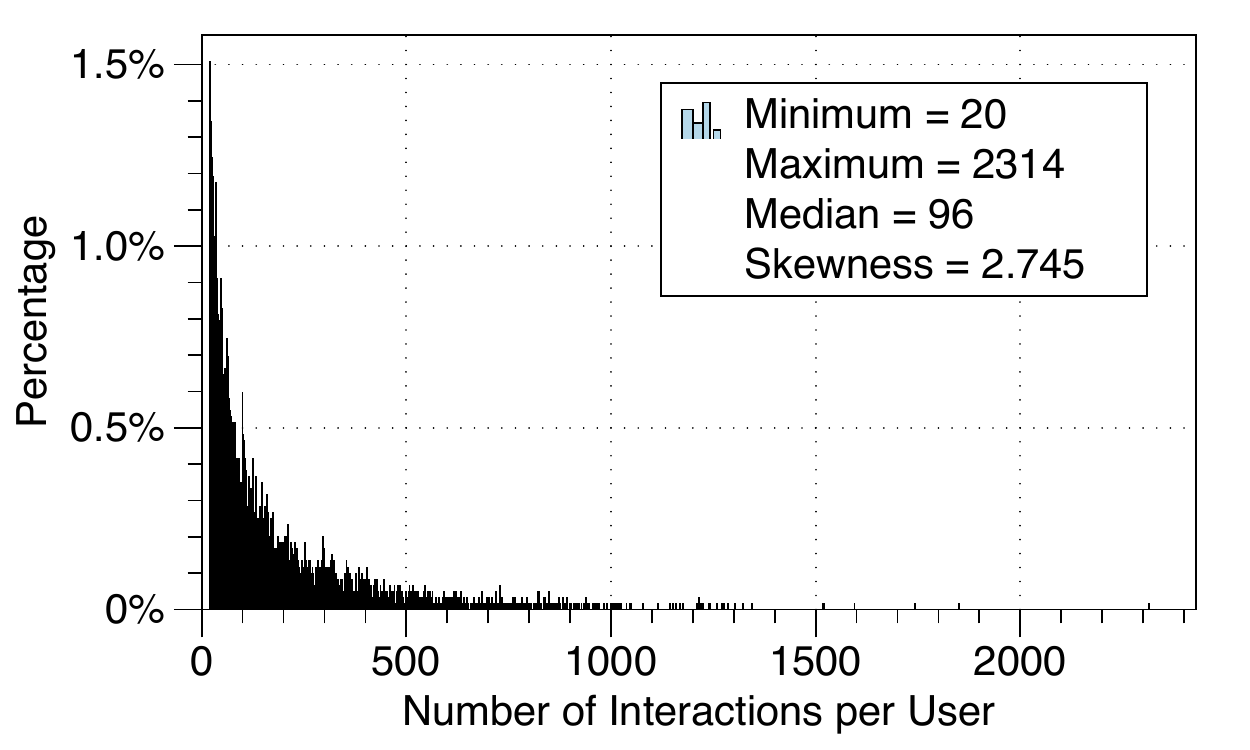}
	\caption{The distribution of the number of interactions per user in the Movielens-1M dataset.}
	\label{fig:uidgb}
\end{figure}

Click-Through Rate (CTR) prediction is an essential task in recommender systems, aiming to predict the probability of a user clicking on a recommended item (e.g., ad, article, product) accurately. Developing deep learning models is becoming the norm to achieve the state-of-the-art CTR prediction performance. 
%Existing deep models~\cite{wide_and_deep,DeepFM,xDeepFM,DBLP:conf/aaai/ChengSH20} typically consist of a module that encodes raw features and captures complex feature interactions, followed by a classifier that predicts the clicking probability. 
%Among the existing deep models~\cite{wide_and_deep,DeepFM,xDeepFM,DBLP:conf/aaai/ChengSH20} for CTR prediction, one important consideration is to learn user preferences from historical interactions in a supervised manner, which is effective for users with sufficient historical behavior data. 
Among the existing deep models~\cite{wide_and_deep,DeepFM,xDeepFM,DBLP:conf/aaai/ChengSH20} for CTR prediction, one important consideration is to learn user preferences from historical interactions, which is effective for users with sufficient interaction data. 
However, most real-world recommender systems involve large numbers of \emph{cold users} who have committed very few interactions, e.g., newly registered users and inactive users.  
As shown in Figure~\ref{fig:uidgb} for example, 20\% of the users with most interactions in the Movielens-1M dataset contribute nearly 60\% of the total interactions. The distribution in practice can be more skewed since Movielens-1M has already filtered out users with fewer than 20 interactions.
With very limited interaction records, the existing deep models for CTR prediction suffer from unsatisfactory prediction performance on cold users~\cite{DBLP:conf/sigir/PanLATH19}. 

%prior efforts to warm-up cold users in ctr prediction
%meta-learning approaches: meta-learner and base learner   + episode training paradigm (two-level nested training)
%- optimization-based：difficult to train meta-learner (gradient over gradient) and need fine-tuning base learner's parameters during test time
% suffer from the need to fine-tune on the target task. 
%- model-based: employ meta-learner to generate base learner's parameters 
%- other approaches:
Great efforts have been devoted to effectively learning preferences of cold users in a broader scope which is known as \emph{user cold-start recommendation}. 
Some researches~\cite{DBLP:conf/aaai/LiJL00H19,DBLP:conf/recsys/BarkanKYK19,DBLP:conf/aaai/XuZCLS20,DBLP:conf/sigir/Hansen0SAL20} focus on completely cold users with no interaction data. In this paper, we consider the more general scenario that cold users have a small number of historical interactions. Recent researches have resorted to developing \emph{meta-learning} algorithms to address the cold-user challenge. The goal of meta-learning is to train a model on a diverse set of tasks, such that the model can learn and adapt quickly to a new task with very few labeled data. 
%Recent researches on the user cold-start challenge have resorted to \emph{meta-learning} algorithms that are able to learn and adapt quickly from very few labeled data.
In the meta-learning framework, cold users are organized into \emph{tasks} (or \emph{episodes}). Each task contains a \emph{support set} involving a few historical interactions of a specific cold user and a \emph{query set} of test items whose interaction labels need to be predicted correctly. 
%[Here maybe refer to figure~\ref{fig:problem}?]
A principled meta-learning algorithm involves two nested learning levels: (i) the \emph{base-learner} works at the level of individual tasks, which acquires user-specific preference information from the support set and perform predictions over the query set; and (ii) the \emph{meta-learner} learns transferable meta-knowledge from different tasks, in order to improve the performance of the base learner across tasks.
Generally, there are two groups of meta-learning algorithms proposed for the user cold-start recommendation.
The first group performs \emph{few-shot user representation learning}. 
They use the meta-learning framework to compute the representation of a particular user by condensing the user's historical interactions within the support set into one fixed-length latent vector via average pooling~\cite{DBLP:conf/nips/VartakTMBL17,DBLP:conf/nips/VolkovsYP17}, attention mechanism~\cite{DBLP:conf/www/LiWW020}, or capsule clustering~\cite{DBLP:conf/sigir/LiangXYY20}. 
%The representation is then used to perform the predictions for the query set. 
{The user representation will be fused with the representation of a query item to perform prediction.} 
However, the condensation can easily cause information loss~\cite{DBLP:conf/ijcai/ZhaoWZ17,DBLP:conf/iccv/PassalisT17,gholamalinezhad2020pooling} and yield suboptimal performance. 
The second group adopts \emph{optimization-based meta-learning}, which is inspired by the MAML algorithm~\cite{DBLP:conf/icml/FinnAL17}.
They focus on the learning of meta-knowledge in the form of good initial values of user ID embeddings~\cite{DBLP:conf/sigir/PanLATH19} or base learner's  parameters~\cite{DBLP:conf/ijcnn/Bharadhwaj19,DBLP:conf/kdd/LeeIJCC19,DBLP:conf/kdd/DongYYXZ20,DBLP:conf/kdd/Lu0S20}. For every task, the base-learner needs to perform fine-tuning using the support set through gradient descent. 
%These approaches are known as \emph{optimization-based meta-learning}, which have two drawbacks. 
However, the optimization-based meta-learning methods have two drawbacks. 
First, the optimization of the meta-learner is expensive in both time and memory cost during training due to the computation of high-order derivatives~\cite{huisman2020survey,hospedales2020meta}. Second, the base learner entails the complexity of fine-tuning when adapting to a new task and thus is inefficient during test time.

To tackle the aforementioned problems, one way is to perform \emph{metric-based meta-learning} (or, metric learning for short) which is a simple yet powerful approach developed for few-shot image classification~\cite{koch2015siamese,DBLP:conf/nips/VinyalsBLKW16,DBLP:conf/nips/SnellSZ17,DBLP:conf/cvpr/SungYZXTH18}. 
The core idea of metric learning is to measure the similarity between a query sample with each of the support samples and infer its label according to the labels of its most similar support samples, which mimics the $k$-nearest neighbor classifier. 
Following typical metric learning methods, the meta-learner involves a feature encoder shared by different tasks that projects raw inputs into a latent feature space, and the base-learner is a classifier that relies on a pre-defined similarity metric such as cosine similarity~\cite{DBLP:conf/nips/SnellSZ17} or a parameterized distance function~\cite{DBLP:conf/cvpr/SungYZXTH18}.

%The key idea is to compute the distance between every query sample with each of the support samples and then predict the label of a query sample according to the label of its nearest support sample. 
%Without loss of generality, the meta-learner in metric-learning involves a feature encoder %shared by different tasks 
%that projects raw inputs into a latent feature space, and the base-learner is a classifier that relies on a  pre-defined similarity metric such as cosine similarity~\cite{DBLP:conf/nips/SnellSZ17} or a parameterized distance function~\cite{DBLP:conf/cvpr/SungYZXTH18}. 
%shenyy: meta-parameter and hyper-parameters, respectively. move to later section??
% (to be optimized through episodic training).  
While metric learning is easy to optimize and avoids test-time adaptation, applying it to few-shot CTR predictions for cold users is still challenging due to the following reason. 
%distinct characteristics of CTR prediction tasks. 
%This is because every few-shot CTR prediction task of a particular user involves two labels, i.e., $1$ for clicking and $0$ for non-clicking, and the two labels are shared by all the tasks. 
That is, in each task, the support set of a particular cold user involves very few interactions that encode limited or even biased information of the user's preference. As a consequence, the base-learner may easily suffer when a query instance is distant from all the instances in the support set in the encoded feature space. 
{{To overcome this limitation, an important observation is that}}
%
%Fortunately, 
different from few-shot image classification where each task contains novel labels to be recognized, every few-shot CTR prediction task of a cold user involves two labels, i.e., one for clicking and zero for non-clicking, and notably the two labels are shared by all the tasks. This \emph{label sharing} among different tasks reveals informative \emph{global preference knowledge}.
%\textcolor{red}{
%\sout{, e.g., popular items are very likely to be clicked by most users, and debased items often receive few clicks.}
%For example, items of some important features such as high score and popular topic are very likely to be clicked by most users, while debased items often receive few clicks.
For example, items with similar features like high ratings and popular categories are very likely to be clicked by most users (i.e., sharing the clicking label), and likewise, debased items often receive few clicks (i.e., sharing the non-clicking label).
Arguably, such global knowledge due to the label sharing fact is useful to complement the sparse and insufficient preference information of cold users (provided by the support sets), and hence has the potential to benefit the CTR prediction performance over the query sets, especially when the query samples are dissimilar or irrelevant to all the support samples.  
It is noteworthy that the
global knowledge is different from the meta-knowledge acquired
by meta-learner as the latter is typically in the form of transferable
embeddings or the initial values of base-learner’s parameters, for the purpose of fast adaptation to new tasks.
To this end, the existing meta-learning approaches mentioned above are incapable of utilizing the global preference knowledge contributed by all the users.
In this paper, we propose \textbf{\modelns} (short for meta-learning \underline{RES}idual \underline{US}er preferences), a generic and efficient approach to address the cold-user challenge in CTR prediction.
The main idea of \model is to decouple user preferences into two parts, namely \emph{basis user preferences} and \emph{residual user preferences}, 
which are learned by different modules and collectively used to predict the probabilities of new clicks. 
%
% The key insight of \model is to decouple the learning of global knowledge on users' preferences from the learning of residual preferences for individual users using different learning algorithms. 
To be specific, \model employs a \emph{shared predictor} to {infer {basis user preferences} on query items based on the input features of query samples.}
%The shared predictor is optimized using the interaction data of different users, and it can be implemented by any CTR prediction architecture~\cite{wide_and_deep,DeepFM,xDeepFM,DBLP:conf/aaai/ChengSH20}.  
The shared predictor can be implemented with any CTR prediction architecture~\cite{wide_and_deep,DeepFM,xDeepFM,DBLP:conf/aaai/ChengSH20}, and {{it is trained with historical interaction data from different users to absorb global preference knowledge.}}
%Specifically, \model employs a \emph{shared predictor} to infer \emph{basis user preferences} on items. 
%The shared predictor can be instantiated by any model proposed for CTR prediction, and is trained using the interaction data of different users in a supervised manner. 
%The predictor can be instantiated by any model proposed for CTR prediction.
%The predictor is used to produce the \emph{basis use preference} on an item, which can be instantiated by any model proposed for CTR prediction.
%
\model then customizes {a \emph{residual preference predictor} as the base-learner to infer \emph{residual user preference} for each query sample according to its matching results with the user-specific historical interactions in the support set.}
%In each task, the meta-learner encodes raw input features via feature embedding and interaction layers, and the base-learner predicts residual user preferences. 
%In \modelns, the meta-learner involves an encoder module that projects input features into a latent feature space and learns high-level feature interactions. The base-learner is trained to infer residual user preferences based on user-specific support sets. 
%Without loss of generality, the meta-learner in metric-learning involves a feature encoder %shared by different tasks 
%that projects raw inputs into a latent feature space, and the base-learner is a classifier that relies on a  pre-defined similarity metric such as cosine similarity~\cite{DBLP:conf/nips/SnellSZ17} or a parameterized distance function~\cite{DBLP:conf/cvpr/SungYZXTH18}. 
%shenyy: meta-parameter and hyper-parameters, respectively. move to later section??
{The rationale of the decoupled preference learning framework is to use the shared predictor to make a rough preference estimation for each query sample based on its input features and then derive residual preference by matching query sample with each of the support samples. To this end, the two components complement each other and \model can perform well when the input features in a query sample are informative to infer its CTR label or the support set is useful to transfer label information to the query sample through matching. According to our experiments, the shared predictor is useful to alleviate the limitation of metric learning when support samples are irrelevant to query samples.}
%
%\textcolor{red}{These two decoupled components are trained with the same training set in a shared feature space but utilize training samples in different ways. Our insight is to first make a rough estimation of basis user preference for each query instance based on its input features, which can be further refined by residual user preference derived from its individual support set. On the other hand, the shared predictor can overcome the aforementioned limitations of typical metric learning. Leveraging the global knowledge shared by collective users, we can counteract a baseless guess when support instances are all distant to the query instance and consequently prevent an arbitrarily bad estimation.
%}
To realize the residual preference predictor (i.e., the base-learner) in \modelns, we provide two efficient designs. %to accomplish the few-shot CTR prediction tasks. 
The first design is a \emph{nearest-neighbor predictor} relying on a similarity function, which is optimized during the training stage without fine-tuning at test time. 
The second design is a \emph{ridge-regression predictor} with differentiable closed-form solvers~\cite{DBLP:conf/iclr/BertinettoHTV19}. It allows task-dependent adaptation during test time, but avoids expensive fine-tuning.  
Certainly, our framework leaves room for other advanced base-learners in the future to be incorporated.
The final CTR prediction result takes the two parts of user preferences into account, which is obtained by fusing the outputs of the shared predictor and the base-learner. 
%the outputs of the shared predictor and the base learner.  
%
We conduct extensive experiments on three public datasets and demonstrate that \model outperforms the state-of-the-art methods in terms of higher CTR prediction accuracy on cold users and lower computational cost. 
%\textcolor{red}{\sout{We further deploy \model in the video recommender system in Wechat Channels where it achieves significant improvement according to the online A/B testing results. }}

%For each support sample, we compute the difference between the given binary label ($1$ or $0$) with the user preference predicted by the shared predictor as the \emph{residual user preference}. 

%to meta-learning \underline{RES}idual \underline{US}er preferences for cold users to improve CTR prediction accuracy.

To summarize, this paper makes the following contributions.
\begin{itemize} 
	
	\item We propose to decouple user preferences into two parts, namely basis user preferences and residual user preferences, and further develop a novel \model approach that employs a shared predictor to capture global preference knowledge to infer basis user preferences and then predicts residual user preferences based on very few user-specific historical interactions. 
%	We propose to fuse the merits of supervised learning and meta-learning to improve CTR prediction accuracy on cold users. We develop a novel \model approach that employs a shared predictor to infer basis user preferences and performs meta-learning to predict residual user preferences based on very few user-specific historical interactions. 
	\item Our proposed \model is a generic decoupled preference learning framework. The shared predictor can be implemented by any model architecture proposed for CTR prediction, and the learning of residual user preferences can be achieved by applying different meta-learning algorithms flexibly. To the best of our knowledge, we are the first to utilize metric-learning for cold-start CTR prediction.
	\item We provide two efficient designs for the base-learner to infer residual user preferences in \modelns: (i) the nearest-neighbor predictor is fast and easy to optimize; and (ii) the ridge-regression predictor performs task-dependent adaptation without entailing the complexity of fine-tuning. 
	\item We conduct extensive experiments on three public datasets demonstrate the superior performance of \model in terms of CTR prediction accuracy on cold users, compared with the state-of-the-art approaches. Further analysis shows that the advantage of \model is more significant for colder users and confirms the efficiency of \model during inference.
	%\textcolor{red}{\sout{ We further deploy \model in a real recommender system and confirm its effectiveness through online A/B testing. }}
\end{itemize}

The remainder of this paper is organized as follows. We present the problem and its meta-learning setting in Section~\ref{sec:prelim}. We describe a basic metric-based meta-learning approach to few-shot CTR prediction in Section~\ref{sec:mus} and elaborate the details of our \model approach in Section~\ref{sec:method}. The experimental results are provided in Section~\ref{sec:evaluation}. We review the related works in Section~\ref{sec:relate} and conclude this paper in Section~\ref{sec:conclude}.

\begin{table}[!t]
	\centering
	\caption{Notation table.}\label{tab:notation}
	%	\resizebox{\linewidth}{!}{
	\begin{tabular}{ll}
	\toprule[1pt]
	Notation&Description\\
	\midrule
	${\bf x}$& Input feature vector in CTR prediction\\
	$y$& Output binary label in CTR prediction\\
	$F$& The number of feature fields\\
	$D$& The whole dataset\\
	%$D_{train}$& The training dataset\\
	$D_u$,$D_u^+$,$D_u^-$&The respective sets of all, positive, and negative\\ &instances of user $u$\\
	$\Theta$& The parameter set of CTR prediction model\\
	$\tau$&The upper limit on the number of interactions \\&of cold users\\
	$U_c$&The set of cold users, $\forall u\in U_c,0<|D_u|\leq \tau$\\
	$T_u$& The few-shot CTR prediction task for user $u$\\
	$S_u$ & The support set in task $T_u$\\
	$Q_u$ &The query set in task $T_u$\\
	$\mathcal{T}_{train,test}$& The train and test data splits in meta-learning setting\\
	%${\bf x}_u^Q, y_u^Q$&input and label of a training instance in $Q_u$\\
	$\Phi$& The feature encoder\\
	$\Psi$& The shared predictor\\
	$\Lambda$& The residual user preference predictor\\
	$\sigma(\cdot)$& The sigmoid function\\
	$g_{\theta}(\cdot,\cdot)$& The similarity function parameterized by $\theta$\\
	$K$& The output dimension of feature encoder $\Phi$\\
	${\bf w}_u$&The task-specific weight vector in \\&ridge regression predictor\\
	$\lambda$&The regularization term in ridge regression predictor\\
	$\beta$&The rescaling coefficient\\
	$\Theta_{meta}$&The set of meta-learner's parameters, i.e., $\Psi,\Phi,\theta,\lambda,\beta$\\
	%$\mathcal{T}_{meta-test}$& the meta-test set
	%of meta-learning setting\\
	%$\mathcal{T}_{meta-validation}$& the meta-validation set
	%of meta-learning setting\\
	
	\bottomrule[1pt]
	\end{tabular}
	%}
\end{table}
	 
\section{Preliminaries}\label{sec:prelim}

In this section, we provide the basic definitions in CTR prediction and then explain how to cast the cold-user CTR prediction problem within the meta-learning framework. Table~\ref{tab:notation} summarizes all the notations used throughout this paper. 

\subsection{Definitions}

%通用的CTR预估问题是，输入为X包括user, item, contextual features，输出为点击行为发生的概率Y，训练数据由已知标签的{X->Y}组成，测试时通过给定X_test预测Y_test；
%%本文方法关注冷用户的CTR预估，即测试的X_test对应的用户u满足|D_train^u|<Threshold，预测其Y_test。

%通用的CTR预估问题是，输入为X包括user, item, contextual features，输出为点击行为发生的概率Y，训练数据由已知标签的{X->Y}组成，测试时通过给定X_test预测Y_test；
%%本文方法关注冷用户的CTR预估，即测试的X_test对应的用户u满足|D_train^u|<Threshold，预测其Y_test。

%[define input to CTR prediction: raw feature space]
Click-Through Rate (CTR) prediction aims to infer the probability that a user would click on a specific item. 
Each \emph{instance} in CTR prediction can be denoted as $({\bf x}, y)$, where ${\bf x}$ is a vector describing $F$ feature fields and $y$ is a binary label indicating a click or non-click behavior. 
In general, the feature fields include user fields (e.g., gender, occupation), item fields (e.g., category, tag), and contextual fields (e.g., time and location when the behavior occurs). 
Typically, ${\bf x}$ is very sparse due to the one-hot encodings of categorical feature fields. We denote by ${\bf e}_i$ the dense embedding of the $i$-th feature field's value in ${\bf x}$. 
Let $D=\{({\bf x},y)\}$ be the set of all the instances.
% and $D_{train}\subseteq D$ be the training set.
%Formally, {CTR prediction} is defined as learning a \emph{binary classifier} $f({\bf x}; \Theta, D_{train})$, where ${\bf x}$ is an input feature vector in the input feature space $\mathcal{X}$, and $\Theta$ denotes the parameter set.  
\begin{definition}[CTR prediction]\label{def:ctr}
	The CTR prediction problem is to train a \emph{binary classifier} $f({\bf x}; \Theta)$, where ${\bf x}$ is an input feature vector in the raw input feature space $\mathcal{X}$, and $\Theta$ denotes the parameter set. 
\end{definition}

%[the collection of data. the collection of user-specific data]
The ultimate goal of this paper is to address the cold-user challenge in CTR prediction. 
%Let $D=\{(x,y)\}$ be all the observed instances.$
Let $D_u\subseteq D$ denote the set of the observed instances associating with user $u$. User $u$ is referred to as a \emph{cold user} if $D_u$ contains a small number of instances. Specifically, we use a \emph{threshold} $\tau$ as the upper limit of instances to identify cold users. Letting $U_c$ be the set of cold users with respect to $\tau$, we have $|D_u|\leq \tau$ for any user $u\in U_c$. 
%The focus of this paper is to learn the preferences of cold users towards higher CTR prediction accuracy.
We assume $|D_u|>0$ for any $u\in U_c$ and leave the zero-shot CTR prediction for completely cold users as future work. 

%[problem: CTR prediction for cold users ] 
\begin{definition}[CTR prediction for cold users]\label{def:problem}
	Consider a cold user $u\in U_c$ with historical CTR instances $D_u$ $(0\leq|D_u|\leq \tau)$. For any feature vector ${\bf x}_u\in \mathcal{X}$ involving $u$, we aim to predict the clicking probability $Pr(y_u =1 \mid {\bf x}_u)$, i.e., $u$'s preference on an item as described in ${\bf x}_u$. 
\end{definition}

\subsection{Meta-learning Framework}
%\subsection{Few-shot Learning Setting}
\begin{figure}[t] 
	\centering
	\includegraphics[width=.8\linewidth]{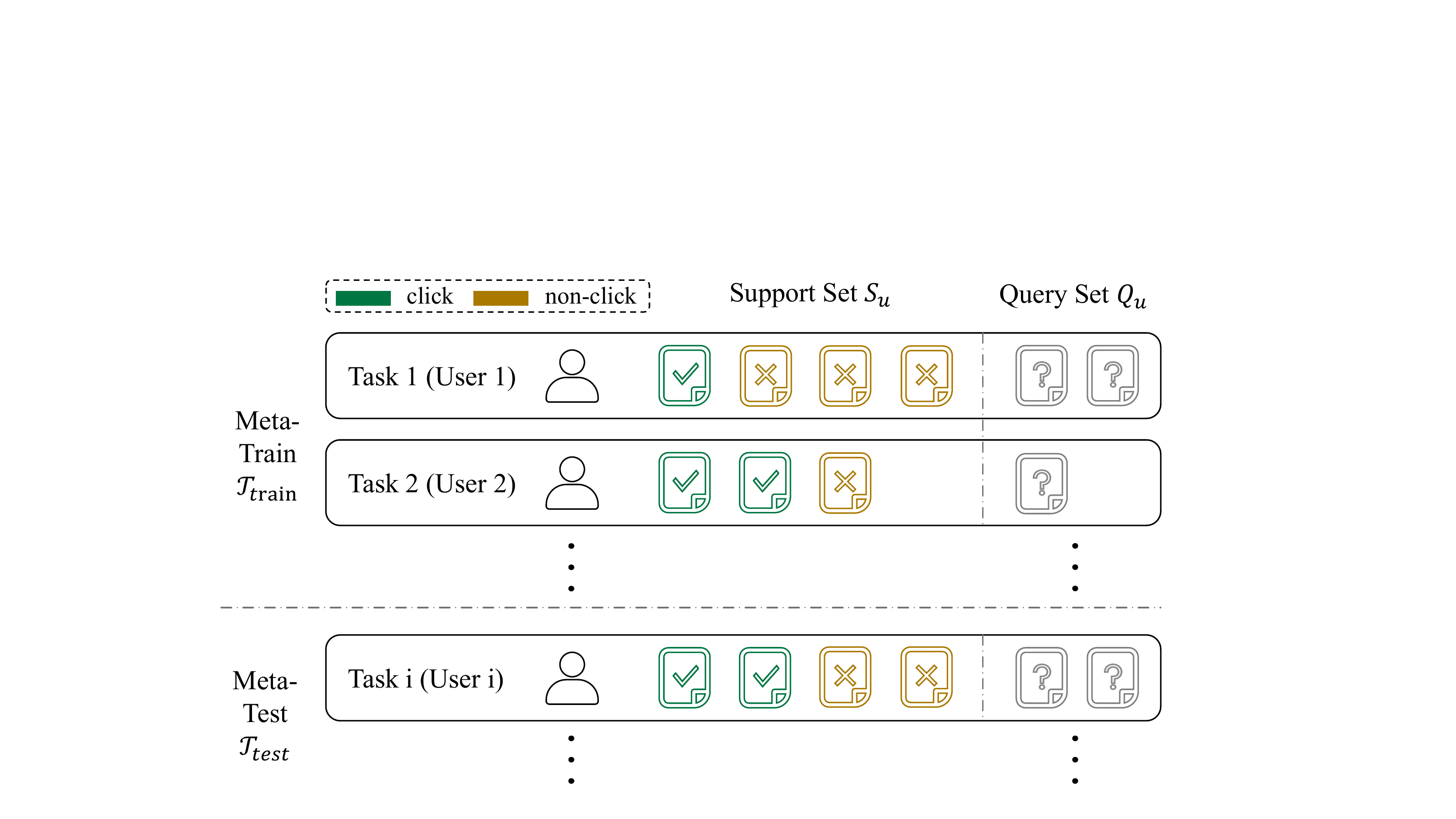}
	\caption{An illustration of the meta-learning framework.}
	\label{fig:problem}
\end{figure}

%Few-shot CTR Prediction Formulation
Learning the preferences of cold users is challenging due to the severely limited information supplied by the user-specific instances $D_u$. Fortunately, performing CTR prediction for cold users bears resemblance to few-shot classification that tries to recognize novel concepts from labeled examples where each label appears only a small number of times. 
% that try to recognize novel classes from very few labeled data.
%The latter tries to propagate label information from a small labeled dataset to unlabeled data.
%In our case, we want to infer cold users' preferences based on very limited historical behavior data. 
\eat{
Specifically, for each cold user $u\in U_c$, we can split $D_u$ into two parts according to the binary labels: $D_u^+$ and $D_u^-$ for $y_u=1$ and $y_u=0$, respectively. 
}
\eat{
In our context, for a cold user $u$, the historical CTR instances $D_u$ are labeled data.
There are two labels in $D_u$ so that we can split $S_u$ into two parts $D_u^+$ and $D_u^-$ according to the label values, i.e.,
$D_u^+=\{({\bf x},y)\in D_u \mid y=1\}$ and $D_u^-=\{({\bf x},y)\in D_u \mid y=0\}$.
Since the preferences on items can vary for different users, we can simply consider the labels to be user-specific. That is, we obtain $2$ labels per cold user and a total of $2|U_c|$ labels with up to $\tau$ instances per label.
%Let $Q_u$ be the set of unlabeled CTR instances associated with $u$, which is referred to as the {\bf query set}.
%Similar to few-shot classification, the goal of CTR prediction is to propagate label information from $S_u$ to $Q_u$, where $|S_u|$ is small.
We now cast CTR predictions for cold users into the following few-shot learning setting, as proposed in~\cite{DBLP:conf/nips/VinyalsBLKW16}.
}
In our context, for a cold user $u\in U_c$, we can split the historical CTR instances $D_u$ into two parts $D_u^+$ and $D_u^-$ according to the label values, i.e.,
$D_u^+=\{({\bf x},y)\in D_u \mid y=1\}$ and $D_u^-=\{({\bf x},y)\in D_u \mid y=0\}$.
Since the preferences on items can vary for different users, here we simply consider the labels to be user-specific. In this way, we can obtain $2$ labels per cold user and a total of $2|U_c|$ labels where each label has up to $\tau$ training instances.
%This is used to adapt the cold start problem setting to traditional FSL settings and  bring in the design of MUS.
Inspired by the similarity of two problems, 
%we cast the problem of learning cold user preference in CTR prediction within the general few-shot problem setting.  Formally, 
%we treat the learning of a cold user's preference as a \emph{few-shot CTR prediction task} or an \emph{episode}.
%we now cast CTR predictions for cold users into the following few-shot learning setting, as proposed in~\cite{DBLP:conf/nips/VinyalsBLKW16}.
we naturally cast CTR prediction for cold users into the standard meta-learning framework which has shown promising performance on few-shot learning problems~\cite{koch2015siamese,DBLP:conf/nips/VinyalsBLKW16,DBLP:conf/nips/SnellSZ17,DBLP:conf/cvpr/SungYZXTH18}.
%following few-shot learning tasks.

Figure~\ref{fig:problem} illustrates the meta-learning framework. We have a {\bf meta-train set} $\mathcal{T}_{train}$ and a {\bf meta-test set} $\mathcal{T}_{test}$ where each set contains a collection of {\bf few-shot CTR prediction tasks} (or tasks for short).
A task $T_u$ aims at promoting CTR prediction performance for cold user $u$. Formally, $T_u$ consists of a {\bf support set} $S_u\subseteq D_u$ and a {\bf query set} $Q_u\subseteq D_u$ that represent the training and test sets for the task respectively, satisfying (i) $|S_u|$ is small and (ii) $S_u\cap Q_u=\emptyset$.
{In general, $S_u$ is formed by randomly selecting labeled instances from $D_u$, and the remaining instances in $D_u$ are randomly sampled to form $Q_u$. When the time information is associated with CTR instances, we shall sort $D_u$ in time order and use the first $|S_u|$ instances to ensure that query instances occur after support instances.} 
%two classes of labeled instances from $D_u^+$ and $D_u^-$, and the remaining instances in $D_u^+$ and $D_u^-$ are randomly sampled to form $Q_u$.
The tasks in $\mathcal{T}_{train}$ and $\mathcal{T}_{test}$ are from two disjoint groups of users, and we defer the details of forming the meta sets in Section~\ref{sec: training} and~\ref{sec:eval_protocol}.
%
%Meta-learning adopts the episode-based training strategy. 
As proposed in~\cite{DBLP:conf/nips/VinyalsBLKW16}, in each training iteration, a task $T_u=(S_u,Q_u)$ is sampled from $\mathcal{T}_{train}$ and a base-learner model is trained to improve the prediction performance over the query set $Q_u$ conditioned on the support set $S_u$. This training procedure matches inference at test time, i.e., predicting clicking probabilities on unlabeled instances for any cold user with a few labeled instances.

While sharing the same meta-learning framework as many few-shot classification problems~\cite{koch2015siamese,DBLP:conf/nips/VinyalsBLKW16,DBLP:conf/nips/SnellSZ17,DBLP:conf/cvpr/SungYZXTH18}, our problem of cold-user CTR prediction has the following distinction. 
The goal of few-shot classification is to acquire the ability of fast adapting to novel concepts where each concept corresponds to a new label. %These novel classes correspond to the label set of the meta-test set. 
In our problem, we actually have two labels, i.e., clicking and non-clicking, shared by all the tasks. 
This label sharing among different tasks (or users) provides insights on global user preference knowledge as described before. 
%The ``novel classes'' correspond to the individual preference of cold users in the meta-test set which are described by the feature vectors ${\bf x}$ rather than the binary click labels. 
%This class sharing inspires us to decouple individual knowledge from global knowledge for modeling cold users' preference.
To clarify the distinction, in what follows, we first provide a basic metric-based meta-learning approach and discuss its limitations. We then present our RESUS approach.

\section{MUS: A Basic Metric-based Meta-learning Approach}\label{sec:mus}
%\section{A Basic Metric-based Approach}

Following the meta-learning framework, for a task $T_u=(S_u, Q_u)$, we aim to learn user $u$'s preference from $S_u$ and utilize it to perform predictions in $Q_u$. A simplest approach to exploit the small support set is to mimic the metric-based meta-learning~\cite{DBLP:conf/nips/VinyalsBLKW16} that relies on a similarity function and transfers label information from support set to query set via matching.
%
%for a task $T_u=(S_u, Q_u)$, a natural way to learn user $u$'s preference from $S_u$ and transfer it to the predictions in $Q_u$ is to mimic the conventional metric-based meta-learning approaches~\cite{DBLP:conf/nips/VinyalsBLKW16}. 
Without loss of generality, metric-based approaches consist of two modules: (1) an \emph{encoder module} as the meta-learner that %projects input feature vectors into a latent feature space;
learns transferrable feature representations for the feature vectors; and (2) a \emph{predictor module} as the base-learner that measures the distances between query and support instances according to their feature representations, and further predicts the label of a query instance using that of its nearest neighbor in the support set.
%In our case, the labels essentially denote the clicking probabilities and hence we predict the label of a query instance by performing a weighted sum over the labels of the support instances using a neural network-based similarity function, instead of referring to the label of its nearest support instance directly. 
Despite its simplicity, the matching strategy is effective for few-shot learning problems and efficient during test time~\cite{DBLP:conf/cvpr/SungYZXTH18}. This motivates us to develop a basic metric-based approach named \textbf{MUS} (\underline{M}eta-learning \underline{US}er preferences) for few-shot CTR prediction. 
%MUS consists of the following two modules.
%a nearest neighbor classifier using a pre-defined metric or a neural network-based similarity function.  
{In the following subsections, we first elaborate on the architecture and training objective of MUS. We then discuss its limitations and provide our insight on developing a decoupled learning framework.}

\subsection{MUS Architecture}
MUS consists of the two modules: the \emph{feature encoder} projects input feature vectors into a latent feature space, and the \emph{user preference predictor} is trained to predict the clicking probabilities of query instances.

\subsubsection{Feature Encoder $\Phi$} 
This module encodes the raw feature vectors of all the instances in $S_u$ and $Q_u$ into dense embedding vectors. This can be realized by any existing CTR prediction model architecture that learns feature embeddings and captures complex feature interactions. By default, we employ the structure of DeepFM~\cite{DeepFM} to implement the feature encoder $\Phi$, which combines the power of factorization machine and deep learning. Recall that we assume $F$ feature fields in each instance. We denote the embedding of the $i$-th feature field in ${\bf x}$ by ${\bf e}_i$, $1\leq i\leq F$. According to DeepFM~\cite{DeepFM}, $\Phi$ is formally defined as follows.
%Various existing CTR prediction model architectures that learn implicit and explicit high-order cross features can be adapted to the structure of encoder $\Phi$. For example, suppose we use DeepFM~\cite{DeepFM} as the encoder architecture, then we have
\begin{align}
%\Phi(x) &= FM(x)\oplus MLP(\mathbf{x}_1\oplus \mathbf{x}_2\oplus...\oplus \mathbf{x}_F)\\
\Phi({\bf x}) &= {\rm FM}({\bf x})\oplus {\rm MLP}({\bf x})\label{eq:encoder1},\\
{\rm FM}({\bf x}) &= (\Sigma_{i=1}^F \mathbf{e}_i)^2-\Sigma_{i=1}^F \mathbf{e}_i^2,\label{eq:encoder2}
\end{align}
where ${\bf x}$ is the feature vector of any support or query instance in $S_u\cup Q_u$. $\oplus$ denotes the concatenation, ${\rm MLP}(\cdot)$ is a multi-layer perceptron.% $F$ is the number of feature fields. 
%\begin{equation}
%FM(x) = (\Sigma_{i=1}^F \mathbf{x}_i)^2-\Sigma_{i=1}^F \mathbf{x}_i^2
%\end{equation}
%Here $\oplus$ denotes vector concatenation, $F$ is the number of input feature fields, and $MLP(\cdot)$ is a multi-layer perceptron.

%\noindent {\bf User Preference Predictor.} 
\subsubsection{User Preference Predictor}
This module is to predict the labels of the query instances based on the support set.
%For ease of description, we use $\Phi({\bf X}_u^S)$ and ${\bf y}_u^S$ to denote the encoded feature vectors and the labels of all the support instances in $S_u$, respectively. Likewise, we have $\Phi({\bf X}_u^Q)$ and ${\bf y}_u^Q$ for the query set $Q_u$, where ${\bf y}_u^Q$ contains the ground-truth labels for prediction.
Note that in our case, the labels essentially denote the clicking probabilities rather than categorical classes. 
Hence, we predict the label of a query instance by performing a weighted sum over the labels of the support instances where the weights are computed based on a similarity function $g_\theta(\cdot,\cdot)$. 
%instead of using the label of its nearest support instance directly. 
Formally, for a query instance $({\bf x}_u^Q, y_u^Q)\in Q_u$, the predictor module computes the clicking probability $\hat{y}_u^Q$ as follows:
\begin{align}
\hat{y}_u^Q &= \sum_{({\bf x}_i^S, y_i^S)\in S_u}{\alpha_i y_{i}^S},\\
\alpha_i &= \frac{\exp\{g_\theta\big(\Phi({\bf x}_u^Q), \Phi({\bf x}_i^S)\big)\}}{\sum_{({\bf x}_j^S, y_j^S)\in S_u} \exp\{g_\theta\big(\Phi({\bf x}_u^Q), \Phi({\bf x}_j^S)\big)\}},
\end{align}
where $g_\theta(\cdot, \cdot)$ can be a neural-network-based function parameterized by $\theta$, or a non-parametric similarity function such as cosine similarity (i.e., $\theta=\emptyset$).  % which can be learned together with the feature encoder. 
%or a neural network-based parameterized function. 

\subsection{Objective Function} 
We use the average cross-entropy as the loss function for the task $T_u$, which is defined as follows:
\begin{equation}
L_{T_u} = -\frac{1}{|Q_u|}\sum_{({\bf x}_u^Q, y_u^Q)\in Q_u} y_u^Q\log\hat{y}_u^Q + (1-y_u^Q)\log(1-\hat{y}_u^Q).
\end{equation}
%During meta-training, the objective is to minimize the loss for all the meta-train tasks. 
%The training objective of MUS on the task $T_u$ is to predict the labels of query instances in $Q_u$ correctly. We employ the average cross-entropy as the loss function, which is defined as follows:
%%the classification loss of the instances in the query set. Considering the CTR prediction task, we have
%\begin{equation}
%L_{u} = -\frac{1}{|Q_u|}\sum_{({\bf x}_u^Q, y_u^Q)\in Q_u} y_u^Q\log\hat{y}_u^Q + (1-y_u^Q)\log(1-\hat{y}_u^Q)
%\end{equation}
During training, the loss $L_{T_u}$ computed on a sampled training task $T_u\in \mathcal{T}_{train}$ is then backpropagated to update the parameters in the feature encoder $\Phi$ and $\theta$ in the predictor.
To be more specific, the parameters in the feature encoder $\Phi$ and $\theta$ are shared across tasks and hence are optimized by minimizing the loss over a batch of training tasks (within the outer loop of meta-learning). 
\subsection{Limitation of MUS and Our Insight}
%\subsection{Discussion on the Limitations}
MUS follows the matching idea as most existing metric-based meta-learning approaches. However, it also inherits the drawback of metric learning. Specifically, due to the small support set, a query instance can be easily distant from all the support instances in the latent feature space, making it difficult to predict the query label accurately. In these cases, the performance of MUS is degenerated to be equivalent to a random guess.

%MUS follows the idea of metric-based meta-learning and treats each user as an independent task. However, as described before, the target labels of different tasks are shared, i.e., clicking or non-clicking. This label dependency across tasks reveals global preference knowledge contributed by collective users. For instance, some popular items would be clicked by most users, while some debased ones are not likely to be clicked by any user. More importantly, due to the small support set, a query instance is easily distant from all the support instances in the latent space, making it difficult to predict the query label accurately without additional knowledge on user preferences.  

To address the limitation, an important observation is that the target labels of different tasks are shared, i.e., clicking and non-clicking. This label sharing across tasks reveals global preference knowledge contributed by collective users. 
As mentioned, some items with high (resp. low) rating scores would be clicked by most (resp. few) users.
%For instance, some \textcolor{red}{\sout{popular }high-score} items would be clicked by most users, while some debased ones are not likely to be clicked by any user. \textcolor{red}{Importance of features such as score and popularity can be revealed by their correlations with labels.}
%
Apparently, exploiting such global preference knowledge has potential benefits to the prediction performance over query sets, especially when the query instances are distant from all the support instances. 
To be more specific, by acquiring global preference knowledge from collective users, we could provide a rough estimation for a query instance of any cold user {based on the input features of the query instance}. {The rough estimation would later be calibrated by referring to user-specific interactions in the support set}. 
%\textcolor{red}{In other words, query instances can play an important role on their own if we can effectively do prediction based on such public knowledge even without referring to the user's support ones.}
However, MUS treats each user as an independent matching task and is thus ignorant of the global preference knowledge among users. This inspires us to decouple the learning of user preferences separately and develop a novel \model framewwork.

\section{The RESUS Approach}\label{sec:method}

In this section, we elaborate the details of our proposed \model approach and the training procedure. 
The key idea of \model is to decouple the predictions of user preferences into two parts. For each query instance in the task $T_u$, we predict \emph{basis user preference} by explicitly exploiting global preference knowledge from the historical interactions of different users. After that, we predict \emph{residual user preference} based on the user-specific information in the support set $S_u$. Finally, we fuse the two preferences to produce the final prediction result.

\begin{figure}[t] 
	\centering
	\includegraphics[width=.9\linewidth]{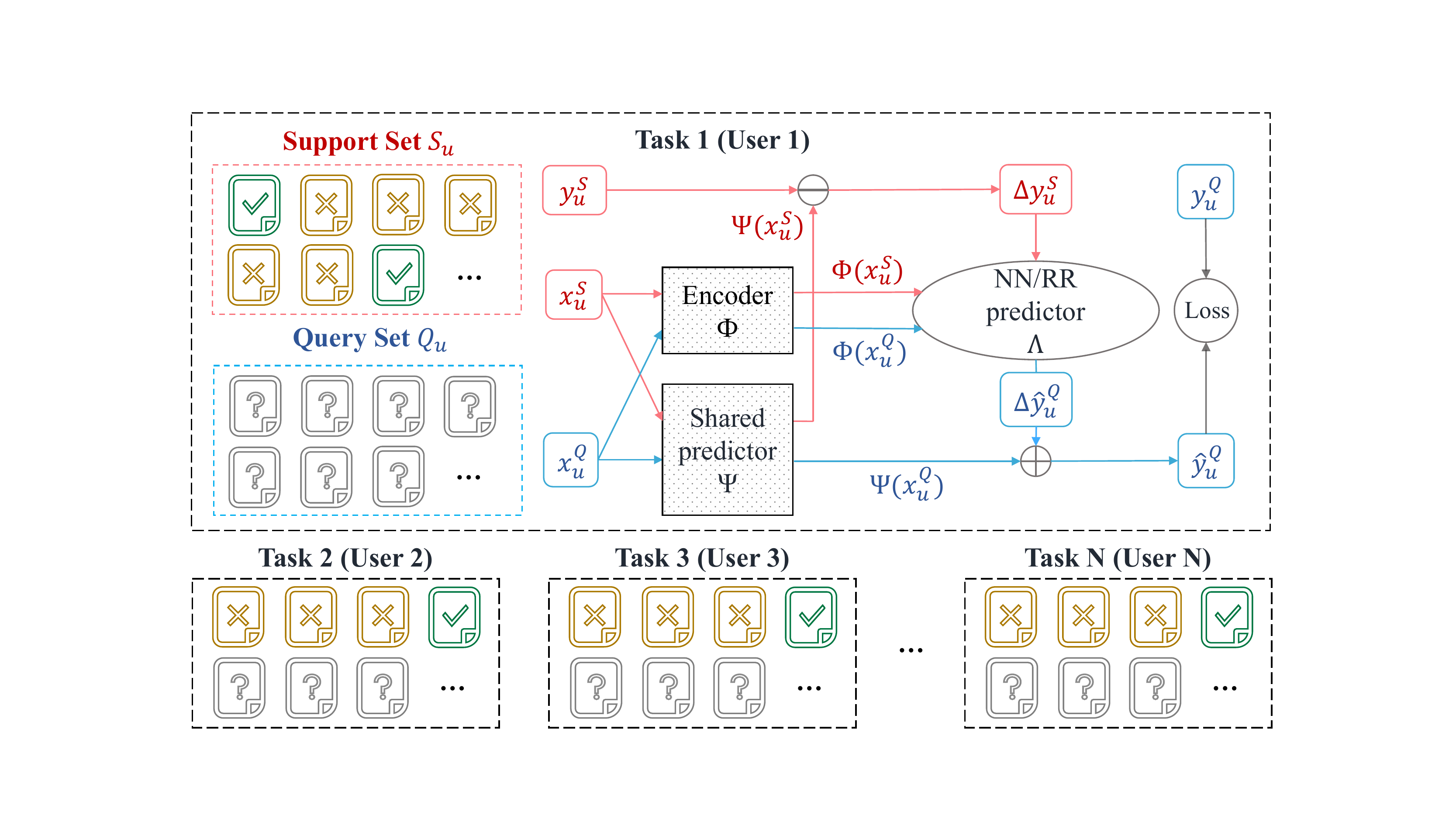}
	\caption{The overview of RESUS. Note that the shared predictor and the encoder are marked by a dotted background, meaning that these two modules are shared across all tasks. In contrast, the NN/RR predictor is a task-specific module.}
	\label{fig:RESUS}
\end{figure}

\subsection{\model Architecture}
Figure~\ref{fig:RESUS} depicts the overview of \model architecture, which consists of three modules: \emph{shared predictor} $\Psi$, \emph{feature encoder} $\Phi$, and \emph{residual user preference predictor} $\Lambda$. 
The feature encoder $\Phi$ follows the same design as the one in MUS using Eq.~(\ref{eq:encoder1})-(\ref{eq:encoder2}).
%For each query instance in the task $T_u$, we accumulate the outputs of two predictors $\Psi$ and $\Lambda$ to produce the final prediction result. 
Therefore, we next present the structures of $\Psi$ and $\Lambda$. Consider the task $T_u=(S_u,Q_u)$.

\subsubsection{Shared Predictor $\Psi$} \label{sec:basis}
This module is to capture global preference knowledge shared by users and predict the basis user preferences for the query instances without referring to the support set. 
The idea is to mimic the existing CTR prediction models that absorb an input feature vector $\mathbf{x}$ and predict the probability of clicking behavior $\hat{y}_{\Psi}$ directly. 
In particular, the implementation of $\Psi$ can be realized by various CTR prediction model structures such as Wide\&Deep~\cite{wide_and_deep}, DeepFM~\cite{DeepFM}, etc. We will evaluate the effects of different structures for $\Psi$ in the experiments (see Table~\ref{tab:base_model}). 
For any instance $({\bf x}, y)$ in $S_u\cup Q_u$, we compute the basis user preference by:
\begin{equation}\label{eq:psi}
\hat y_\Psi=\sigma(\Psi({\bf x})),
%\hat y_\Psi=\Psi({\bf x})
\end{equation}
%$\hat y_\Psi=\sigma(\Psi({\bf x}))$ as the basis user preference, 
where $\sigma(\cdot)$ is the sigmoid function to control the predicted results within the range of $(0,1)$.
% $\hat y_\Psi=\sigma(\Psi({\bf x}))$ as the basis user preference, where $\sigma(\cdot)$ is the sigmoid function. 

In \modelns, $\Psi$ is trained across tasks using the historical interactions of different users (see Section~\ref{sec: training}). In this way, it is able to acquire global preference knowledge from collective users and based on which, it can preliminarily assess the user preference according to the input feature vector. 
 
%Intuitively, $\Psi$ assess the user preference solely based on the feature vector ${\bf x}$. 

\subsubsection{Residual User Preference Predictor $\Lambda$}

In this module, we first compute the residual user preference for each support instance $({\bf x}_u^S, y_u^S)\in S_u$ as follows:
\begin{equation}\label{eq:residual}
\Delta {y}_u^S = {y}_u^S-\hat{y}_{\Psi}^S, %; or  \Delta {y}_u^S = {y}_u^S-\sigma(\Psi({\bf x}_u^S)) ??
\end{equation}
where $\hat{y}_{\Psi}^S = \sigma(\Psi({\bf x}_u^S))$. 
Henceforth, we obtain the transformed support set $S'_u=\{ ({\bf x}_u^S, \Delta{y}_u^S) \}$.

%shenyy:till here
We now focus on predicting the residual user preferences for query instances based on the transformed user-specific training data $S'_u=\{ ({\bf x}_u^S, \Delta{y}_u^S) \}$.
%For ease of description, we use $\Phi({\bf X}_u^S)\in \mathbb{R}^{|S_u|\times K}$ ($K$ is the dimension of encoded feature vectors) and ${\bf y}_u^S\in \mathbb{R}^{|S_u|\times 1}$ to respectively denote the encoded feature vectors and the corresponding labels of all the support instances in $S_u$, stacked as rows. Likewise, we have $\Phi({\bf X}_u^Q)$ and ${\bf Y}_u^Q$ for the query set $Q_u$. %, where ${\bf y}_u^Q$ contains the ground-truth labels for prediction.
%In this module, we first compute the residual user preferences for the support instances as follows:
%\begin{equation}
%\Delta {\bf Y}_u^S = {\bf Y}_u^S-\sigma(\Psi({\bf X}_u^S))
%\end{equation}
%%
%%We then encode the input feature vectors ${\bf X}_u^S$ and ${\bf X}_u^Q$ into the latent vectors $\Phi({\bf X}_u^S)$ and $\Phi({\bf X}_u^Q)$ using the feature encoder, respectively. 
%%
%By transforming the support set $S_u$ into $\big(\Phi({\bf X}_u^S), \Delta{\bf Y}_u^S \big)$, we now focus on predicting the residual user preferences for $\Phi({\bf X}_u^Q)$ based on the user-specific training data $\big(\Phi({\bf X}_u^S), \Delta{\bf Y}_u^S\big)$.
We propose two efficient ways to implement the residual user preference predictor $\Lambda$: (i) the nearest-neighbor predictor is a similarity-based regression model, which prevents fine-tuning on new tasks; and (ii) the ridge-regression predictor enables fast adaptation to new tasks using the closed form solver. 
It is worth mentioning that both predictors are fully differentiable, enabling end-to-end optimization of \modelns.

\vspace{.03in}
{\bf (i) {Nearest-neighbor (NN) predictor.}} This is similar to the user preference predictor in MUS. 
It makes the prediction of a query instance's residual preference as a weighted sum over the residual preferences of the support samples.  
%where the weights are computed using a fixed or learned similarity measure.
Formally, for a query instance $({\bf x}_u^Q, y_u^Q)\in Q_u$, the NN predictor computes the residual user preference $\Delta{\hat{y}_u^Q}$ as follows: 
\begin{align}
	\Delta\hat{y}_u^Q &= \sum_{({\bf x}_i^S, \Delta y_i^S)\in S_u'}{\alpha_i \Delta y_{i}^S},\label{eq:predict1}\\
	\alpha_i &= \frac{\exp\{g_\theta\big(\Phi({\bf x}_u^Q), \Phi({\bf x}_i^S)\big)\}}{\sum_{({\bf x}_j^S, \Delta y_j^S)\in S_u'} \exp\{g_\theta\big(\Phi({\bf x}_u^Q), \Phi({\bf x}_j^S)\big)\}}.
\end{align}
Similar to MUS, $g_\theta(\cdot,\cdot)$ is a similarity function, which can be non-parametric such as cosine similarity or a neural network parameterized by $\theta$.
In \modelns, we implement $g$ as follows:
\begin{equation}
g_{\theta}(\mathbf{v}_1, \mathbf{v}_2) = \mathbf{w}^\top\lVert \mathbf{v}_1-\mathbf{v}_2\rVert+b,
\end{equation}
where $\mathbf{w}\in \mathbb{R}^{K \times 1}$, $K$ is the output dimension of feature encoder $\Phi(\cdot)$, and $b$ is a bias term.
Similar to MUS, $\theta=\{{\bf w},b\}$ is shared among tasks and hence can be viewed as the hyperparameters of the predictor $\Psi$. In particular, $\theta$ is optimized over training tasks and is fixed during test time.

\vspace{.03in}
{\bf (ii) {Ridge-regression (RR) predictor.}} 
This is a task-specific predictor that trains its parameters based on the support set in each task. 
To avoid expensive training from scratch or fine-tuning over the support set per task, we employ ridge regression that admits a closed form solution~\cite{DBLP:conf/iclr/BertinettoHTV19} that can be computed directly in the inner loop of meta-learning.
%Consider a task $T_u=(S_u,Q_u)$.
 For ease of description, we use $\Phi({\bf X}_u^S)\in \mathbb{R}^{|S_u|\times K}$ and $\Delta{\bf y}_u^S\in \mathbb{R}^{|S_u|\times 1}$ to respectively denote the encoded feature vectors and the residual preferences of all the support instances in $S_u$, stacked as rows.

The ridge regression is parameterized by ${\bf w}_u\in\mathbb{R}^{K\times 1}$ and fit by solving the following optimization problem over $S_u$:
\begin{equation}\label{eq:ridge}
\underset{\mathbf{w}_u}{\rm minimize}\left\Vert \Delta {\bf y}_u^S-\Phi({\bf X}_u^S)\mathbf{w}_u\right\Vert ^{2}+\lambda\left\Vert \mathbf{w}_u\right\Vert ^{2},
\end{equation}
where $\lambda\geq 0$ is optimized within the outer loop of meta-learning.
The closed form solution for Eq.~(\ref{eq:ridge}) is the following: 
\begin{equation}\label{eq:rr}
{\bf w}_u^*=\big(\Phi({\bf X}_u^S)^\top\Phi({\bf X}_u^S)+\lambda I\big)^{-1}\Phi({\bf X}_u^S)^\top\Delta {\bf y}_u^S.
\end{equation}
Note that the computation of Eq.~(\ref{eq:rr}) involves an inversion operation over the $K\times K$ matrix. To alleviate the expensive computation cost, we adopt the Woodbury formula~\cite{petersen2008matrix} to obtain ${\bf w}_u^*$ as follows:
\begin{equation}\label{equ:wood}
\mathbf{w}_u^*=\Phi({\bf X}_u^S)^\top\big(\Phi({\bf X}_u^S)\Phi({\bf X}_u^S)^\top+\lambda I\big)^{-1}\Delta {\bf y}_u^S.
\end{equation}
The inversion is now performed over the $|S_u|\times|S_u|$ matrix. Since $|S_u|\ll K$ and $|S_u|$ is typically very small (due to cold users), we can reduce the cost of computing ${\bf w}_u^*$ significantly.

Given ${\bf w}_u^*$, the RR predictor computes the residual user preference for a query instance $({\bf x}_u^Q, y_u^Q)\in Q_u$ as follows:
\begin{equation}\label{eq:predict2}
\Delta \hat{y}_u^Q = \Phi({\bf x}_u^Q)^\top\mathbf{w}_u^*.
\end{equation}
%\textcolor{red}{
%Note that the signature of $\Delta \hat{y}_u^Q$ is related to $\Delta {\bf y}_u^S$ in both NN and RR predictor, which follows our intuition.
%}
\emph{Remark.}
The predicted residual preference $\Delta \hat{y}_u^Q$ using Eq.~(\ref{eq:predict1}) or Eq.~(\ref{eq:predict2}) is dependent on $\Delta {\bf y}_u^S$ and may yield negative values.

\subsubsection{Putting Two Kinds of Preferences Together}

By far, for each query instance $({\bf x}_u^Q, y_u^Q)$ in $Q_u$, we obtain the predicted basis user preference $\hat y_\Psi^Q=\sigma(\Psi({\bf x}_u^Q))$ using Eq.~(\ref{eq:psi}), and infer the residual user preference $\Delta {\hat y}_u^Q$ using Eq.~(\ref{eq:predict1}) or Eq.~(\ref{eq:predict2}). 
We then fuse them together to produce the final CTR prediction result $\hat{y}_u^Q$ in the following way.
\begin{equation}\label{eq:fuse}
%\hat{y}_u^Q = \sigma(\beta\Delta{\hat y}_u^Q + \hat{y}_\Psi)
\hat{y}_u^Q = \sigma(\Psi({\bf x}_u^Q)+\beta\Delta{\hat y}_u^Q ).
\end{equation}
%where $\sigma(\cdot)$ is the sigmoid function. 
Note that we fuse $\Delta{\hat y}_u^Q$ with $\Psi({\bf x}_u^Q)$, followed by the sigmoid function to normalize the final prediction within the range of $(0,1)$. $\beta$ is a hyperparameter of the base learner and used as the rescaling coefficient for the calibration purpose. 
Intuitively, when the input feature vector involves discriminative information to determine the preference score, we expect the predicted basis user preference to be close to the ground-truth $y_u^Q$ even if the base-learner suffers from insufficient user-specific preference information in the support set. Likewise, when the shared predictor can only infer a rough preference score according to the general preferences among users, we encourage the residual user preference to refine the rough score based on the user-specific historical interactions. 
In our work, $\beta$ is treated as a parameter of the meta-learner and optimized over training tasks in the outer-loop of meta-learning. As an alternative, one may set the value of $\beta$ in a more fine-grained way, e.g., $\beta$ is shared among tasks with the same support set size $|S_u|$. The intuition is that users with more historical interactions can be assigned with a large value of $\beta$.

\subsection{Training Procedure}\label{sec: training}

\begin{algorithm}[t!]
	\SetAlgoLined
	\SetNlSty{}{}{:}
%	\ShowLn\KwIn{The meta-train set $\mathcal{T}_{train}$.}
	\ShowLn\KwIn{The set of training users $U_{train}$ with $D_{train}$.}
	\ShowLn\KwOut{The meta-learner parameters $\Theta_{meta}$.}
%\tcc{$\Theta_{meta}=\{\Psi,\Phi,\theta,\lambda,\beta\}$}
Randomly initialize $\Theta_{meta}$\;
%Pretrain $\Psi$ using loss defined in  Equation~(\ref{equ:pretrain}).
\tcc{Train $\Psi$}
\While{not converged}{
	Sample a batch of instances $\mathcal{B} $ from $D_{train}$\;
	Evaluate $L_{\Psi}$ over $\mathcal{B}$ based on Eq.~(\ref{equ:pretrain})\;
	Update the parameters in $\Psi$ by minimizing $L_{\Psi}$\;
}
Freeze the parameters in $\Psi$\;
\tcc{Train $\Theta_{meta}\backslash\Psi$}
\While{not converged}{
%给训练集每个用户采样support和query set
%为每个用户计算loss，并反向传播梯度
%达到batch size后，进行更新
%\For{\texttt{$u \in \mathcal{T}_{train}$}}
Sample a batch of training users $\mathcal{B}_u$ from $U_{train}$\;
\For{$u \in \mathcal{B}_u$}{
%        \State{
    \uIf{$\mathcal{P}_{|S_u|}$ is available}{
    Sample support set size $|S_u|\sim \mathcal{P}_{|S_u|}$\;}
    \Else{
    Sample support set size $|S_u|\sim \mathcal{U} \{1,\tau\}$\;
    } 
	Get $S_u$ by sampling $|S_u|$ instances from $D_u$\;
	Get query set $Q_u=D_u\backslash S_u$\;
	Evaluate $L_{T_u}$ based on Eq.~(\ref{equ:lu})\;
	}
	Update $\Theta_{meta}\backslash\Psi$ by minimizing $\frac{\Sigma_{T_u \in \mathcal{B}_T} |Q_u|L_{T_u}}{\sum_{T_u\in \mathcal{B}_T}{|Q_u|}}$\;
%    \If{$|D_u|>|S_u|$}{
%	Sample $S_u$, $Q_u$ from $D_u$\;}}}
}
\caption{RESUS - meta-learning RESidual USer preferences}\label{alg:optimize}
\end{algorithm}

Before presenting the training procedure for \modelns, we first describe the construction of the \emph{meta-train set} $\mathcal{T}_{train}$. We construct $\mathcal{T}_{train}$ by sampling {historical} behaviors of a set of training users $U_{train}$. % on a per epoch basis. 
{Specifically, for each training user $u\in U_{train}$ with historical behaviors $D_u$, we randomly select $|S_u|$ instances from $D_u$ to form the support set and use the remaining instances as the query set, i.e., $Q_u=D_u\backslash S_u$.  
In our experiments, when the time of interactions is known, we use the $|S_u|$ instances with the smallest timestamps rather than random sampling to preserve the time order between support set and query set. }
%
%$u$'s clicking \textcolor{red}{\sout{instances from $D_u^+$ }and non-clicking instances from \sout{$D_u^-$} $D_u$} to form the support set $S_u$, and leave the remaining instances as the query set, i.e., $Q_u=D_u\backslash S_u$.
%\textcolor{red}{Specially, if the behaviors have time information, we choose the earliest $|S_u|$ instances instead of random ones to make sure all query instances occur after support instances.}
%
The size of the support set $|S_u|$ is sampled from the distribution $\mathcal{P}_{|S_u|}$ of the observed interaction numbers of actual cold users in real-world recommender systems, {which is defined as follows:
\begin{equation}
	P(|S_u|=i) = \frac{\sum_{u'\in U_{cold}} \mathbb I(|D_{u'}|=i)}{|U_{cold}|},
\end{equation}
where $i\in\{1,\cdots, \tau\}$, $U_{cold}$ is the set of cold users and $\mathbb I$ is the indicator function. Recall that $\tau$ is the upper limit on historical instances to identify cold users.
It is desirable to sample support set size from the actual distribution to mitigate the mismatch between training and inference tasks. In the case where the distribution of the number of interactions on cold users is unknown, we sample the size of support set $|S_u|$ from a uniform distribution $\mathcal{U} \{1,\tau\}$, i.e., $P(|S_u|=i)=1/\tau$ for $i\in\{1, \cdots, \tau\}$.  In the experiments, as public datasets save filtered out users with very few numbers of historical interactions, we adopt the uniform distribution to sample the size of support set.}
The above sampling process to form $\mathcal{T}_{train}$ is repeated at the beginning of each epoch.

%“" 

\eat{
During training, for each user, we randomly sample her historical click behaviors $S_u^+$ and non-click behaviors $S_u^-$ into the support set and leave the remaining ones as the query set. 
The size of the support set $|S_u|$ is also sampled from the distribution of testing cold users $\mathcal{P}_{test}(|S_u|)$ or a uniform distribution $\mathcal{U} \{1,\tau\}$ if $\mathcal{P}_{test}(|S_u|)$ is not available. Recall that $\tau$ is the threshold of recognizing cold users.
This sampling process is repeated at the beginning of each epoch.
}

In \modelns, only ${\bf w}_u$ (in ridge-regression predictor) is learned within each task (with the closed-form solution), and all the other parameters are optimized over training tasks within the outer loop of meta-learning, including: 
(i) the parameters in the shared predictor $\Psi$;
(ii) the parameters in the feature encoder module $\Phi$;
(iii) the hyperparameters in the residual user preference predictor $\Lambda$, i.e., $\theta$ or $\lambda$;
(iv) the rescaling coefficient $\beta$ for fusing two preferences in Eq.~(\ref{eq:fuse}).
For ease of description, we use $\Theta_{meta}$ to denote the set of parameters in  (i)-(iv). %=\{\Psi,\Phi,\theta,\lambda,\beta\}$.
For a training task $T_u=(S_u,Q_u)\in \mathcal{T}_{train}$, 
%$\Theta_{meta}$ is optimized by minimizing the binary cross entropy: 
we compute the binary cross entropy loss as follows:
%\begin{equation}
%	L = -\frac{1}{|Q_u|}\sum_{i=1}^{|Q_u|} Y_{Q_i}log\hat{Y}_{Q_i} + (1-Y_{Q_i})log(1-\hat{Y}_{Q_i})
%\end{equation}
\begin{equation}\label{equ:lu}
%L_{u} = -\frac{1}{|Q_u|}\sum_{({\bf x}_u^Q, y_u^Q)\in Q_u} y_u^Q\log\hat{y}_u^Q + (1-y_u^Q)\log(1-\hat{y}_u^Q)
L_{T_u} = -\frac{1}{|Q_u|}\sum_{({\bf x}_u^Q, y_u^Q)\in Q_u} y_u^Q\log\hat{y}_u^Q + (1-y_u^Q)\log(1-\hat{y}_u^Q),
\end{equation}
where $\hat{y}_u^Q$ is the predicted clicking probability computed by Eq.~(\ref{eq:fuse}). %The total loss of \model is then defined by:
Hence, $\Theta_{meta}$ is optimized by minimizing the following loss function over $\mathcal{T}_{train}$:
\begin{equation}\label{eq:loss}
%L_{RESUS} = \frac{1}{\sum_{T_u \in \mathcal{T}_{train}} |Q_u|}\sum_{T_u \in \mathcal{T}_{train}} L_{T_u}
L_{RESUS} = \frac{\sum_{T_u=(S_u,Q_u) \in \mathcal{T}_{train}} |Q_u|L_{T_u}}{\sum_{T_u=(S_u,Q_u) \in \mathcal{T}_{train}} |Q_u|}.
\end{equation}
%\begin{align} 
%L_{RESUS} &= \frac{1}{\sum_{u \in \mathcal{T}_{train}} |Q_u|}\sum_{u \in \mathcal{T}_{train}} |Q_u|L_{u}\\
%&=\frac{1}{\sum_{u \in \mathcal{T}_{train}} |Q_u|}-\sum_{u \in \mathcal{T}}\sum_{({\bf x}_u^Q, y_u^Q)\in Q_u} y_u^Q\log\hat{y}_u^Q + (1-y_u^Q)\log(1-\hat{y}_u^Q)
%\end{align}
%
The above loss function is a weighted sum of $L_{T_u}$, $T_u \in \mathcal{T}_{train}$. In practice, we use batch gradient descent and accumulate the gradients of a batch of training tasks to make an update to the parameters in $\Theta_{meta}$.

While the parameters in the shared predictor $\Psi$ can be optimized through meta-learning, we found it is more beneficial to train $\Psi$ in advance and fix it during the whole meta-learning process. 
Specifically, we {use all the} historical interactions {from training users $U_{train}$ to pretrain $\Psi$}. Let $D_{train}=\{({\bf x}, y)\in D_u  \mid u\in U_{train} \}$ denote the interactions of the training users{, where $D_u=S_u\cup Q_u$}. 
%
%Specifically, we first use all the instances in meta-train set $\mathcal{T}_{train}$ to pretrain $\Psi$. Let $D_{train}=\{({\bf x}, y)\in S_u\cup Q_u \mid T_u=(S_u, Q_u)\in \mathcal{T}_{train} \}$. 
%
We learn $\Psi$ by minimizing the following loss function:
% [the support sets and query sets in meta-train set? Both of them are sampled from meta-train set, so no need to say so?]
\begin{equation}\label{equ:pretrain}
%L_{\Psi} = -\frac{1}{|\mathcal{T}_{train}|}\sum_{({\bf x}, y)\in \mathcal{T}_{train}} y\log\hat{y}_{\Psi} + (1-y)\log(1-\hat{y}_{\Psi})
%L_{\Psi} = -\frac{1}{|\mathcal{T}_{train}|}\sum_{T_u\in \mathcal{T}_{train}} \sum_{({\bf x}, y)\in S_u\cup Q_u} y\log\hat{y}_{\Psi} + (1-y)\log(1-\hat{y}_{\Psi})
L_{\Psi} = -\frac{\sum_{({\bf x}, y)\in D_{train}} y\log\hat{y}_{\Psi} + (1-y)\log(1-\hat{y}_{\Psi})}{|D_{train}|},
\end{equation}
where $\hat{y}_{\Psi}$ is the predicted basis user preference using Eq.~(\ref{eq:psi}). In practice, we update the paremeters in $\Psi$ using batch gradient descent.
After pretraining, we freeze the parameters of $\Psi$ during the optimization of the other parameters in RESUS. 
Intuitively, this training setting prevents interference between the learning of basis user preferences and residual user preferences, empirically leading to faster convergence and better performance (see the results in Table~\ref{tab:hyper1}). 
%This prevents  interference between $\Psi$ and the other modules in \model during meta-learning, empirically leading to faster convergence and better performance (see the results in Figure~\ref{fig:hyper}). 
The overall training procedure of \model is summarized in Algorithm~\ref{alg:optimize}.

%Besides, we found that for the predictor $\Psi$ in the meta-learner, a pre-training and then fixed manner is beneficial to the final performance of RESUS. Specifically, we first use the labeled data of training users and the support set of testing users to train the predictor $\Psi$. We then froze the parameters of $\Psi$ during the training of RESUS. This helps to alleviate interference between the base learner and the predictor during training and empirically leads to faster convergence and better performance.

%}

\subsection{Discussions}\label{sec:discuss}

%\textcolor{red}{
\subsubsection{Comparison with MUS}
{It is easy to verify that our basic metric-based meta-learning approach {MUS} described in Section~\ref{sec:mus} is equivalent to \model (using the NN predictor) without the shared predictor $\Psi$ (i.e., always predicting zero)}. 
Recall that MUS suffers from the sparse information in the support sets of cold users. In \modelns, the shared predictor $\Psi$ can be instantiated with any existing CTR models.
%to learn global preference knowledge from collective users. 
It provides rough inference results for query instances based on the input features and the predictions are not affected by the utility of support sets. The predicted basis user preferences, though may not be accurate, are of great assistance in the final prediction performance.
Nevertheless, \model is not a simple combination of the shared predictor and MUS. Specifically, the metric-learning counterpart in \model is not independent of the shared predictor but utilizes it in two ways. First, metric-learning relies on the shared predictor to encode the feature vectors of support and query instances. As we pretrain the shared predictor $\Psi$ in a fully supervised manner, it can gain insights from interactions of collective users to derive good representations of feature vectors that are beneficial to the matching performance.
Second, thanks to the shared predictor, the tasks performed by metric-learning in RESUS focus on fitting residual user preferences. Similar to the idea of boosting, inferring residuals is relatively simpler than inferring the overall preferences.
Our experimental results also confirm the advantages of the decoupled preference learning framework, compared with MUS (see Table~\ref{tab:hyper1}) and different shared predictors (see Table~\ref{tab:base_model}). 
 {It is also important to notice that MUS and \model are trained using the same training set $D_{train}$. Particularly, the shared predictor in \model is pretrained over $D_{train}$ in a supervised manner (in order to acquire global preference knowledge from different users) and the metric-learning counterpart (i.e., $\Theta_{meta}$)  is optimized with the support and query sets sampled from $D_{train}$ on an individual user basis (with the purpose of fast acquiring individual preference knowledge from  support set). }

%In MUS, chances are that the query instances cannot match any irrelevant support instances, and the support set helps little even if the feature encoder $\Phi$ can perform well on the support set. While in our \model, the shared predictor $\Psi$ utilizes global knowledge such as side information which is not subject to the cold start problem. Then we can approximately and even precisely obtain accurate user preferences from the query instances alone. Consequently, the basis user preference $\Psi({\bf x}_u^Q)$ works as a reliable complement to the residual user preference $\Delta{\hat y}_u^Q$, in case $\Delta{\hat y}_u^Q$ falls into a random guess.
%}

\subsubsection{Time Complexity Analysis}

To predict the clicking probability of a query instance for user $u$, the time complexity of \model is determined by three modules: the shared predictor $\Psi$, the feature encoder $\Phi$, and the residual user preference predictor $\Lambda$. 
%[Here we consider the computation for a single query set sample.]
First, since the shared predictor is applied to each sample in the support set $S_u$, the time complexity of $\Psi$ is ${\rm O}(|S_u||W_{\Psi}|)$, where $|W_{\Psi}|$ is the number of parameters in $\Psi$. Now that we can use any typical deep CTR prediction model architecture (e.g., DeepFM~\cite{DeepFM}) to implement $\Psi$, $W_{\Psi}$ refers to the set of weights in the corresponding neural network.
Second, the time complexity of $\Phi$ is ${\rm O}(|S_u||W_{\Phi}|)$.
Third, we have two choices for the residual user preference predictor $\Lambda$. The time complexity of the NN predictor is ${\rm O}(|S_u|K)$, and that of the RR predictor is ${\rm O}(|S_u|^2K+|S_u|^3)$, where $K$ is the dimension of the encoded feature vectors. 
Finally, the total time complexity of \modelns$_{NN}$ is ${\rm O}(|S_u|(|W_{\Psi}|+|W_{\Phi}|+K))$, and that of \modelns$_{RR}$ is ${\rm O}(|S_u|(|W_{\Psi}|+|W_{\Phi}|+|S_u|K+|S_u|^2))$.
Note that $|S_u|$ and $K$ are much smaller than $|W_{\Psi}|+|W_{\Phi}|$, since $u$ is a cold user and $S_u$ is usually from tens to hundreds. Hence the time complexity of \model is ${\rm O}(|S_u|(|W_{\Psi}|+|W_{\Phi}|)$.

%Besides, it is worth noticing that the above time complexity analysis is based on a single sample.
For a batch of query samples, the time complexity of \model becomes ${\rm O}(B|S_u|(|W_{\Psi}|+|W_{\Phi}|))$, where $B$ is the batch size.
However, \model can be implemented in a user-based batch manner where the query samples of the same user correspond to one batch and the results of $\Psi$ and $\Phi$ are computed once within the batch. In this way, the batch time complexity of \model can be reduced to ${\rm O}((B+|S_u|)(|W_{\Psi}|+|W_{\Phi}|))$.

In comparison, the existing meta-learning approaches for few-shot CTR prediction on cold users can be generally divided into two groups: few-shot user representation learning approaches (e.g., NLBA~\cite{DBLP:conf/nips/VartakTMBL17}) and optimization-based meta-learning approaches (e.g., MeLU~\cite{DBLP:conf/kdd/LeeIJCC19}). 
The batch time complexities of the two groups of approaches are ${\rm O}((B+|S_u|)|W_{\Phi}|)$ and ${\rm O}((B+|S_u|)|W_{\Psi}|)$, respectively.
%The batch time complexity of existing few-shot user representation learning methods (e.g., NLBA~\cite{DBLP:conf/nips/VartakTMBL17}) and optimization-based meta-learning methods (e.g., MeLU~\cite{DBLP:conf/kdd/LeeIJCC19}) are typically $O((B+|S_u|)|W_{\Phi}|)$ and $O((B+|S_u|)|W_{\Psi}|)$, respectively. 
Since $|W_{\Phi}|$ and $|W_{\Psi}|$ are typically close, the time complexity of \model is the same as the existing meta-learning methods. 
However, since optimization-based meta-learning approaches need to perform error backpropagation to update the base-learner for every new task, they are empirically more time-consuming during test time (see the results in Table~\ref{tab:time}). %than few-shot user representation learning methods and \modelns. Experimental comparison are shown in Section~\ref{sec:time_cost}.

%$O(|S_u|(|W_{\Psi}|+|W_{\Phi}|+|S_u|K+|S_u|^2)+B(|W_{\Psi}|+|W_{\Phi}|+|S_u|K))$.

%$O((|S_u|+1)*(|W_{\Psi}|+|W_{\Phi}|))$
%
%NN: $|S_u|K$
%
%NN total: $O(|S_u|(|W_{\Psi}|+|W_{\Phi}|))+O(|S_u|K)$
%
%RR: $|S_u|^2K+|S_u|^3$
%
%RR total: $O(|S_u|(|W_{\Psi}|+|W_{\Phi}|))+O(|S_u|^2K+|S_u|^3)$
%user batch: $O(|S_u|(|W_{\Psi}|+|W_{\Phi}|+|S_u|K+|S_u|^2)+B(|W_{\Psi}|+|W_{\Phi}|+|S_u|K))$
%
%mixed batch: $O(B|S_u|(|W_{\Psi}|+|W_{\Phi}|+|S_u|K+|S_u|^2)$
%
%meta batch: $O(|S_u||W_{\Psi}|+B|W_{\Psi}|)$

\subsubsection{Comparison with Gradient Boosting}

%Boosting is an ensemble machine learning algorithm for converting weak learners to strong ones~\cite{zhou2019ensemble}.
Gradient boosting~\cite{friedman2001greedy} is a powerful ensemble technique that tries to convert weak learners to a strong one~\cite{zhou2019ensemble}. The idea of gradient boosting is to train multiple learners in an iterative fashion. Each learner attempts to fit the errors of its predecessor, and all the learners are then combined to give the final prediction results. Here we highlight the key differences between \model and gradient boosting.
First and foremost, in \modelns, the residual user preference predictor used for one user-specific task can be viewed as an individual learner. %(though the residual predictors for different tasks can share parameters like the NN predictor). 
Each learner tries to fit the residual preferences for a particular user on the query instances. This means different learners do not collaborate with each other. 
% to perform predictions for all the users. 
In gradient boosting, however, the predictions from all the learners are accumulated to deliver one output for an input instance. 

Second, while both the shared predictor $\Psi$ and the residual user preference predictor $\Lambda$ contribute to the final prediction of an unlabeled instance, they are utilized for different purposes: $\Psi$ learns the association from input feature vectors to output binary labels by exploiting global preference knowledge from collective users; 
$\Lambda$ refers to the support set to infer the user-specific residual preferences. 
In contrast, gradient boosting ensembles multiple weak learners with the same purpose, i.e., fitting the residual errors to form a strong learner.

Third, gradient boosting performs optimization on the additive term and hence the learners are iteratively optimized. 
In contrast, \model is cast in the typical meta-learning framework and the parameters $\Theta_{meta}$ is optimized via the end-to-end task-based training procedure.  
%In \modelns, the hyperparameters in the RR predictor is trained in the outer loop of meta-learning over training tasks and the parameters of the NN predictor is optimized with the closed-form solver, which allows us to perform an end-to-end meta-learning of \modelns. 
%
%First, while the residual user preference predictor $\Lambda$ learns the residual error of the shared predictor $\Psi$, this error is defined in a user-specific way in order to improve the prediction accuracy for a particular user, instead of being shared by all the users. [In GB, all the users share the same model, so the residual error are summed up among different users for optimization.]
%
%Second, gradient boosting aims to turn multiple weak learners (e.g., decision trees) into a strong one. While in \modelns, the shared predictor and the residual predictor are employed for different purposes, i.e., capture global preference knowledge and learn user-specific residual preferences, respectively.
%
%Third, in gradient boosting, each learner is iteratively optimized through gradient to fit the residual errors of its predecessor. 
%In \modelns, we use nearest-neighbor or ridge regression predictor to learn the residual errors, which are both closed-form solvers that allows end-to-end optimization of the whole model. 
%
We also conduct the comparison experiment with gradient boosting in Section~\ref{sec:compare_boosting}.

\section{Experiments}
\label{sec:evaluation}

In this section, we conduct experiments to answer the following research questions:

\begin{enumerate}[label=\textbf{RQ\arabic*:}]
	\item How does our proposed \model approach perform on cold users compared with the state-of-the-art methods?
%	\item What are the effects of residual module, predictor pre-training, and the choice of base model architecture in our proposed method?
	\item How do different deigns of the key components in \model affect its performance?
%	How do the key components of \model affect its performance? %(i.e., residual module, predictor pre-training, choice of base model architecture) 
	\item How does \model perform compared with gradient boosting method?
	\item	{How does \model perform when support instances are irrelevant to query instances?}
	\item	What is the empirical time cost of \modelns?
	%How does \model perform on the empirical time cost compared with other methods?
\end{enumerate}

%In what follows, we first describe the experimental settings, followed by results answering the above three questions.

\subsection{Experimental Settings}

%\begin{table}[t]
%	%	\captionsetup{labelfont=bf}
%	\centering
%%	\small
%	\caption{The statistics of experimental datasets.}
%	\label{tab:datasets}
%	\begin{tabular}{l|c|c|c|c}
%		\toprule[1pt]
%		\textbf{Datasets}&\textbf{\#Users}&\textbf{\#Items}
%%		&\textbf{\makecell{\#Feature\\Fields}}
%		&\textbf{\#FeatureFields}
%		&\textbf{\#Records}
%%		&\textbf{Features}
%		\\
%		\midrule[0.5pt] 
%		Movielens&6,040&3,706&7&1,000,209\\
%		Frappe&957&4,082&10&288,609\\
%		Taobao&1,141,729&846,811&16&26,557,961\\
%		\bottomrule[1pt]
%	\end{tabular} 
%\end{table}

%\begin{table}[t]
%	%	\captionsetup{labelfont=bf}
%	\centering
%	\small
%	\caption{The statistics and feature fields of two datasets.}
%	\label{tab:datasets}
%	\begin{tabular}{c|c|c}
%		\toprule[1pt]
%		&\textbf{Movielens}&\textbf{Frappe}\\
%		\midrule[0.5pt] 
%		\#users&6,040&957\\
%		\#items&3,706&4,082\\
%		\#records&1,000,209&288,609\\
%		%		Number of features&9,788&5,382\\
%		Sparsity&95.53\%&92.61\%\\
%		\midrule[0.5pt] 
%		\makecell{Feature\\fields}&\makecell{Gender, Age, Genre, \\Occupation, Release year }&\makecell{Country, City, Daytime, \\Weekday, Weekend,\\Location, Cost, Weather}\\
%		\bottomrule[1pt]
%	\end{tabular} 
%\end{table}

\subsubsection{Datasets}
We experiment with three publicly available datasets. The statistics of the three datasets are summarized in Table~\ref{tab:datasets}. Note that some widely-used CTR prediction datasets like Criteo~\cite{Criteo} and Avazu~\cite{Avazu} are not applicable in our experiments since the records are anonymous and cannot be organized into user-specific tasks. The details of the experimental datasets are as follows.
%their features are anonymous without user identification.
\begin{itemize}
\item {\bf Movielens}~\cite{Movielens}: This is the one million version of the Movielens dataset, which consists of users' ratings on movies. We convert the ratings to binary labels by setting a threshold $3$, i.e., ratings $\geq 3$ and ratings $< 3$ are marked as $1$ and $0$, respectively. It contains feature fields about users (i.e., user ID, age, gender and occupation) and about items (i.e., movie ID, genre and release year). 
\item {\bf Frappe}~\cite{frappe15}: This dataset contains app usage logs from users under different contexts {(e.g., weekday, location) without timestamp information}. We converted each log 
%(user features, app features, context information) 
to a feature vector as input. The target value indicates whether the user has used the app in the context.

\item {\bf Taobao}~\cite{DIN}: This dataset contains ad display/click logs from an e-commercial website. It involves both user profiles (e.g., gender, age, and occupation) and item features (e.g., category, brand, price). The target value indicates whether the user has clicked the advertisement.
\end{itemize}
%{\bf $\bullet$ Movielens}~\cite{Movielens}: This is the one million version of the Movielens dataset, which consists of users' ratings on movies. We convert the ratings to binary labels by setting a threshold $3$, i.e., ratings $\geq 3$ and ratings $< 3$ are marked as $1$ and $0$, respectively. It contains feature fields about users (i.e., user ID, age, gender and occupation) and about items (i.e., movie ID, genre and release year). 

%{\bf $\bullet$ Frappe}~\cite{frappe15}: This dataset contains app usage logs from users under different contexts (e.g., daytime, location). We converted each log (user features, app features, context information) to a feature vector as input. The target value indicates whether the user has used the app in the context.
%
%{\bf $\bullet$ Taobao}~\cite{DIN}: This dataset contains ad display/click logs from an e-commercial website. It contains both user profiles (e.g., age, occupation, consumption grade) and item features (e.g., category, brand, price). The target value indicates whether the user has clicked the advertisement.

For all the above datasets, we filter out items with fewer than $100$ interactions during preprocessing to alleviate the effects from cold items, since we focus on the cold-user issue.
\begin{table}[t]
	%	\captionsetup{labelfont=bf}
	\centering
%	\small
	\caption{The statistics of the datasets.}
	\label{tab:datasets}
	\begin{tabular}{lccc}
		\toprule[1pt]
		&\textbf{Movielens}&\textbf{Frappe}&\textbf{Taobao}\\
		\midrule[0.5pt] 
		\#Users&6,040&957&1,141,729\\
		\#Items&3,706&4,082&846,811\\
		Sparsity&95.53\%&92.61\%&99.99\%\\
		\#Feature Fields&7&10&16\\
		\#Features&9,789&5,382&3,436,646\\
		%		Number of features&9,788&5,382\\
		\#Instances&1,000,209&288,609&26,557,961\\
		%		\midrule[0.5pt] 
		%		\makecell{Feature\\fields}&\makecell{Gender, Age, Genre, \\Occupation, Release year }&\makecell{Country, City, Daytime, \\Weekday, Weekend,\\Location, Cost, Weather}\\
		\bottomrule[1pt]
	\end{tabular} 
\end{table}
\subsubsection{Evaluation Protocols}\label{sec:eval_protocol}

%In our implementation, we form a meta-train set for each epoch. 
%To form the meta-training set $\mathcal{T}_{train}$, we first sample a set of tasks (i.e., users). For each task of user $u$, we randomly select $P_u^+$ clicking behaviors and $N_u^-$ non-clicking behaviors of $u$ to compose the support set, and use the remaining behavior data of $u$ as the query set.
To evaluate CTR prediction performance on cold users, for all the datasets, we first randomly split them into training, validation and test sets by user ID with the ratio of $7:2:1$. This ensures that the users in different data splits are non-overlapped.
%During evaluation, 
{For Movielens and Taobao datasets, we sort each user's interactions in time order.} For each user in the validation or test set, we {use the first} $|S_u|$ instances from the user's historical interactions $D_u$ to form the support set and leave the remaining ones as the query set for testing. {As Frappe does not include timestamp information, we randomly sample $|S_u|$ instances from $D_u$ to form the support set for a test user $u$ and use the remaining samples as the query set.}
 
To evaluate model performance on users with different degrees of coldness, we control the value of $|S_u|$ in meta-test tasks. At each time, we set a uniform value of $|S_u|$ for all the test users {which is no greater than the upper limit $\tau$ (we set $\tau$ to be 30, 30 and 150 for Movielens, Frappe and Taobao, respectively)}. 
%Note that to evaluate model performance on cold users with different training instances, each time, we set a uniform value of $|S_u|$ for all the users and then change the value for the next time. 
%
For Movielens and Frappe datasets, $|S_u|$ is chosen from $\{1,2,\cdots,30\}$. For Taobao dataset, since the dataset is unbalanced (i.e., most instances are negative), we run experiments for $|S_u|$ in $\{10,20,\cdots,150\}$. 
To better present the results, we divide each of the value sets equally into three cold-start stages: {\bf Cold Start-\uppercase\expandafter{\romannumeral1}/\uppercase\expandafter{\romannumeral2}/\uppercase\expandafter{\romannumeral3}} and report the averaged test performance of all the approaches in each stage. Specifically, the three stages for Movielens and Frappe correspond to $|S_u|$ in $\{1,2,\cdots,10\}$, $\{11,12,\cdots,20\}$, $\{21,22,\cdots,30\}$, respectively. The three stages for Taobao correspond to $|S_u|$ in $\{10,20,\cdots,50\}$, $\{60,70,\cdots,100\}$, $\{110,120,\cdots,150\}$, respectively.
We use Logloss and AUC as the metrics, 
which are widely-used for evaluating CTR prediction methods. Following previous work~\cite{DIN,DBLP:conf/icml/YanLXH14}, we further use $RelaImpr$ to measure the relative improvement of AUC, as defined:
\begin{equation}
RelaImpr = \big(\frac{\text{AUC(target model)}-0.5}{\text{AUC(base model)}-0.5}-1\big)\times 100\%.	
\end{equation}
%\begin{equation}
%RelaImpr = \big(\frac{\text{AUC/GAUC(target model)}-0.5}{\text{AUC/GAUC(base model)}-0.5}-1\big)\times 100\%
%\end{equation} 
We ran each experiment for $10$ times and reported the averaged results.

\subsubsection{Comparison Methods} 
We consider two kinds of meta-learning approaches as the comparison methods: 
%We compare our proposed methods with the following representative baselines, including both few-shot user representation learning approaches and optimization-based meta-learning approaches.

\vspace{.03in}
\noindent{\underline{\emph{(1) Few-shot user representation learning methods:}}}

\begin{itemize}
\item \textbf{LWA}~\cite{DBLP:conf/nips/VartakTMBL17}. This method encodes a user's interaction  records (in the support set) into latent vectors and computes two embeddings by applying average pooling over the user's positive and negative latent vectors, respectively. {It then computes a weighted sum over two embeddings to obtain the user representation and encodes all the features in a query instance into a latent vector as the item representation. The user and item representations are concatenated to predict the preference for the query instance via a linear classifier.}  
\item \textbf{NLBA}~\cite{DBLP:conf/nips/VartakTMBL17}. This method is similar to LWA, but uses a non-linear classifier to model the interactions between user and item representations.
\item \textbf{JTCN}~\cite{DBLP:conf/sigir/LiangXYY20}. This method first {encodes representations of historical items with the dynamic routing-by-agreement mechanism of capsule networks, and then applies attentive aggregation to generate a fixed-length high-level user preference representation, which is later fused with embeddings of user features to output user representation. The predictions are made based on the user representation and the embeddings of query items.}
\end{itemize}
{All the above three methods optimize model parameters via gradient descent, especially optimizing the encoders to generate user and item representations. They take both historical user behaviors in support set and the input features in query instance as input features to each CTR prediction. However, they encode historical user behaviors into one fixed-length latent vector, which may suffer from information loss. Note that in our \model approach, each historical user behavior in a support set contributes to the prediction of a query instance separately.}

%\noindent $\bullet$ \textbf{LWA}~\cite{DBLP:conf/nips/VartakTMBL17}. 
%	This method encodes a user's interaction  records (in the support set) into latent vectors and computes two embeddings by applying average pooling over the user's positive and negative latent vectors, respectively. It then computes a weighted sum over two embeddings to obtain the user representation, which will be used to predict the preferences for the items in the query set. 
%
%\noindent $\bullet$ \textbf{NLBA}~\cite{DBLP:conf/nips/VartakTMBL17}. This method is similar to LWA, but uses a non-linear classifier to model the interactions between user and item representations.
%
%\noindent $\bullet$ \textbf{JTCN}~\cite{DBLP:conf/sigir/LiangXYY20}. This method first computes high-level user interests from historical interaction data with the dynamic routing-by-agreement mechanism of capsule networks, and then applies attentive aggregation to generate fixed-length user representation. The predictions are made based on the user and item representations. 
\vspace{.03in}
\noindent\underline{\emph{(2) Optimization-based meta-learning methods:}}
\begin{itemize}
\item \textbf{Meta-Embedding}~\cite{DBLP:conf/sigir/PanLATH19}. This is a state-of-the-art MAML-like method. It trains an item ID embedding generator based on item content features~\cite{DBLP:conf/icml/FinnAL17}, so that the ID embeddings of cold items fine-tuned on a small number of support samples can perform well on the query set. In our experiments, we employ Meta-Embedding for cold users for comparison.

\item \textbf{MeLU}~\cite{DBLP:conf/kdd/LeeIJCC19}. This is another MAML-like method, which learns an initialization scheme for the base learner's parameters for fast adaption on a small number of historical interactions of cold users. 
%It  also uses the meta-optimization algorithm of MAML~\cite{DBLP:conf/icml/FinnAL17} for optimizing the model. 

\item \textbf{MAMO}~\cite{DBLP:conf/kdd/DongYYXZ20}. This method {is another state-of-the-art method which} improves MeLU with memory-augmented networks. It designs task-specific memory and feature-specific memory to guide the model with personalized network parameter initialization.
\end{itemize}
{Note that the optimization-based meta-learning methods typically consume high training time and memory cost due to the bi-level optimization. Their goal is to generate a good initialization for base-learner’s parameters. The parameters will be fine-tuned through gradient updates based on the support set. Therefore, while fine-tuning is beneficial to the predictions of query instances, it hurts inference time. In contrast, the base-learners (i.e., NN and RR predictors) in RESUS do not require gradient updates during test. }
%These methods assume that support set can provide query set with potential optimal parameters according to gradients of loss on support set, while error predictions of support instances in RESUS are downplayed via matching mechanism.}

%\noindent $\bullet$ \textbf{Meta-Embedding}~\cite{DBLP:conf/sigir/PanLATH19}. This is a state-of-the-art MAML-like method. It trains an item ID embedding generator based on item content features~\cite{DBLP:conf/icml/FinnAL17}, so that the ID embeddings of cold items fine-tuned on a small number of support samples can perform well on the query set. In our experiments, we employ Meta-Embedding for cold users for comparison.
%
%\noindent $\bullet$ \textbf{MeLU}~\cite{DBLP:conf/kdd/LeeIJCC19}. This is another MAML-like method, which learns an initialization scheme for the base learner's parameters for fast adaption on a small number of historical interactions of cold users. It  also uses the meta-optimization algorithm of MAML~\cite{DBLP:conf/icml/FinnAL17} for optimizing the model. 
%
%\noindent $\bullet$ \textbf{MAMO}~\cite{DBLP:conf/kdd/DongYYXZ20}. This method improves MeLU with memory-augmented networks. It designs task-specific memories and feature-specific memories to guide the model with personalized network parameter initialization.
%%\end{itemize}

\vspace{.03in}
\noindent\underline{\emph{Our proposed methods:}}
\begin{itemize}
\item \textbf{RESUS$_{NN}$}. This is our \model approach using the nearest-neighbor predictor as the base-learner. 

\item \textbf{RESUS$_{RR}$}. This is our \model approach using the ridge-regression predictor as the base-learner.%, which has a differentiable closed-form solution. 
\end{itemize}

%%\begin{itemize}
%\noindent $\bullet$ \textbf{RESUS$_{NN}$}. This is our \model approach using the nearest-neighbor predictor as the base-learner. %which is a non-parametric model.
%
%\noindent $\bullet$ \textbf{RESUS$_{RR}$}. This is our \model approach using the ridge-regression predictor as the base-learner.%, which has a differentiable closed-form solution. 
%%\end{itemize}

Note that the basic metric-based meta-learning approach {\bf MUS} described in Section~\ref{sec:mus} is equivalent to \modelns$_{NN}$ without the shared predictor $\Psi$ (i.e., always predicting zero). Hence, we report the results of MUS when evaluating the effects of $\Psi$ in Section~\ref{sec:rq2}.

\subsubsection{Implementation Details}

We implemented our proposed methods based on Pytorch\footnote{Our code is available at \url{https://github.com/WeiyuCheng/RESUS}}. %~\cite{DBLP:conf/nips/PaszkeGMLBCKLGA19}. 
By default, we used DeepFM~\cite{DeepFM} as the architecture of the shared predictor $\Psi$ and that of the feature encoder $\Phi$ (excluding the final prediction layer). 
%We used DeepFM~\cite{DeepFM} which is a typical and widely-used CTR prediction model, as the architecture of encoder and predictor in RESUS by default. 
For LWA, NLBA, JTCN, Meta-Embedding, MeLU and MAMO, we employed the same DeepFM architecture as the interaction function for a fair comparison. 
{For each dataset, Meta-Embedding includes all the available feature fields in its CTR instances. RESUS and the other baselines use all the feature fields except user ID as they use user ID to form meta-tasks. Following the original paper of Meta-Embedding, we pretrain its base model using $D_{train}$ including user ID.} 
We
% employed Xavier initialization~\cite{pmlr-v9-glorot10a} for all the model parameters and 
applied Adam optimizer~\cite{DBLP:journals/corr/KingmaB14} with an initial learning rate of $0.001$ and a mini-batch size of $1024$ by default. We set the dimension of input feature embeddings to $10$ for all the comparison methods. To avoid overfitting, we performed early-stopping according to model's AUC on the validation set. We also tuned the hyperparameters of the baselines using the validation set. All the experiments were conducted on a Linux server equipped with Intel Xeon 2.10GHz CPUs and NVIDIA GeForce RTX 2080Ti GPUs. %We will make our code publicly available upon paper acceptance.

\subsection{Performance Comparison (RQ1)} 

\begin{table*}[ht]
	\centering
	\caption{Performance comparison on three datasets, where the bold values are the best results and the underlined values are the two most competitive results (RQ1). {* indicates a statistically significant level $p$-value<0.05 comparing \model with the best baseline.}}\label{tab:main_result}
  \resizebox{\textwidth}{!}{
	\begin{tabular}{cllccccccccc}
		%{cl|l|ccc|ccc|ccc}
		\toprule
		\multirow{10}{*}{\begin{sideways}\bf Movielens\end{sideways}} &
		\multirow{2}{*}{\bf Method Class}\!\!\!\! &  
		\multirow{2}{*}{\bf Method} & \multicolumn{3}{c}{\bf Cold Start-\uppercase\expandafter{\romannumeral1}} & \multicolumn{3}{c}{\bf Cold Start-\uppercase\expandafter{\romannumeral2}} & \multicolumn{3}{c}{\bf Cold Start-\uppercase\expandafter{\romannumeral3}} \\
		&    &   & \bf Logloss & \bf AUC   &\bf  RelaImpr & \bf Logloss& \bf AUC   & \bf RelaImpr &\bf Logloss& \bf AUC   & \bf RelaImpr \\
		\cmidrule{2-12}
		&\multirow{3}{*}{\makecell[l]{Few-Shot User\\ Representation Learning 
		}}\!\!\!\!
		&LWA&0.3982 &0.7216 &0.0\% &0.3718 &0.7498 &0.0\% &0.3765 &0.7510 &0.0\% \\
		&&NLBA&0.3824 &0.7302 &3.9\% &0.3693 &0.7528 &1.2\% &0.3739 &0.7542 &1.3\% \\
		&&JTCN&0.3371 &0.7518 &13.6\% &0.3400 &0.7494 &-0.2\% &0.3474 &0.7476 &-1.4\% \\
		\cmidrule{2-12}
		&\multirow{3}{*}{\makecell[l]{Optimization-Based \\Meta-Learning}} 
		&MeLU&0.3342 &0.7574 &16.1\% &0.3309 &0.7689 &7.7\% &0.3340 &0.7707 &7.8\% \\
		&&MAMO&\underline{0.3342} &{0.7575} &{16.2\%} &\underline{0.3300} &\underline{0.7691} &\underline{7.7\%} &\underline{0.3325} &\underline{0.7710} &\underline{8.0\%} \\
		&&Meta-Embedding\!\!\!\!&0.3351 &\underline{0.7577} &\underline{16.3\%} &0.3347 &0.7625 &5.1\% &0.3390 &0.7655 &5.8\%\\
		\cmidrule{2-12}
		&
		\multirow{2}{*}{Proposed Methods 
		}&RESUS$_{NN}$&\underline{0.3322} &\underline{0.7638}$^*$ &\underline{19.0\%} &\textbf{0.3274}$^*$ &\underline{0.7745}$^*$ &\underline{9.9\%} &\textbf{0.3306}$^*$ &\underline{0.7771}$^*$ &\underline{10.4\%} \\
		&&RESUS$_{RR}$&\textbf{0.3318} &\textbf{0.7645}$^*$ &\textbf{19.3\%} &\underline{0.3280}$^*$ &\textbf{0.7772}$^*$ &\textbf{11.0\%} &\underline{0.3314}$^*$ &\textbf{0.7793}$^*$ &\textbf{11.3\%} \\
		\midrule
		\multirow{10}{*}{\begin{sideways}\bf Frappe\end{sideways}} &
		\multirow{2}{*}{\bf Method Class}\!\!\!\! &  
		\multirow{2}{*}{\bf Method} & \multicolumn{3}{c}{\bf Cold Start-\uppercase\expandafter{\romannumeral1}} & \multicolumn{3}{c}{\bf Cold Start-\uppercase\expandafter{\romannumeral2}} & \multicolumn{3}{c}{\bf Cold Start-\uppercase\expandafter{\romannumeral3}} \\
		&    &   & \bf Logloss & \bf AUC   &\bf  RelaImpr & \bf Logloss& \bf AUC   & \bf RelaImpr &\bf Logloss& \bf AUC   & \bf RelaImpr \\
		\cmidrule{2-12}
		&\multirow{3}{*}{\makecell[l]{Few-Shot User\\ Representation Learning 
		}}\!\!\!\!
		&LWA&0.3696 & 0.7732 & 0.0\% & 0.3334 & 0.8098 & 0.0\% & 0.3112 & 0.8302 & 0.0\% \\
		&&NLBA&0.3751 & 0.8430 & 25.6\% & 0.3594 & 0.8435 & 10.9\% & 0.3713 & 0.8440 & 4.2\% \\
		&&JTCN&\underline{0.3130} & 0.8306 & 21.0\% & 0.3178 & 0.8248 & 4.8\% & 0.3167 & 0.8277 & -0.8\% \\
		\cmidrule{2-12}
		&\multirow{3}{*}{\makecell[l]{Optimization-Based \\Meta-Learning}} 
		&MeLU &0.3241 &0.8459 &26.6\% &\underline{0.3084} &0.8520 &13.6\% &0.2916 &0.8596 &8.9\% \\
		&&MAMO&\textbf{0.3055} &0.8340 &22.3\% &\underline{0.2896} &0.8535 &14.1\% &\underline{0.2867} &0.8674 &11.3\% \\
		&&Meta-Embedding\!\!\!\!&\underline{0.3128} &\textbf{0.8640} &\textbf{33.3\%} &0.2914 &\underline{0.8663} &\underline{18.2\%} &\underline{0.2714} &\underline{0.8715} &\underline{12.5\%} \\
		\cmidrule{2-12}
		&
		\multirow{2}{*}{Proposed Methods 
		}&RESUS$_{NN}$&  0.3322 & \underline{0.8542} & \underline{29.7\%} & \textbf{0.2842}$^*$ & \textbf{0.8746}$^*$ & \textbf{20.9\%} & \textbf{0.2561}$^*$ & \textbf{0.8931}$^*$ & \textbf{19.0\%} \\
		&&RESUS$_{RR}$& 0.3275 & \underline{0.8524} & \underline{29.0\%} & 0.3106 & \underline{0.8669} & \underline{18.4\%} & 0.2896 & \underline{0.8713} & \underline{12.4\%}\\

		\midrule
		\multirow{10}{*}{\begin{sideways}\bf Taobao\end{sideways}} &
		\multirow{2}{*}{\bf Method Class}\!\!\!\! &  
		\multirow{2}{*}{\bf Method} & \multicolumn{3}{c}{\bf Cold Start-\uppercase\expandafter{\romannumeral1}} & \multicolumn{3}{c}{\bf Cold Start-\uppercase\expandafter{\romannumeral2}} & \multicolumn{3}{c}{\bf Cold Start-\uppercase\expandafter{\romannumeral3}} \\
		&    &   & \bf Logloss & \bf AUC   &\bf  RelaImpr &\bf  Logloss& \bf AUC   & \bf RelaImpr &\bf Logloss& \bf AUC   & \bf RelaImpr \\
		\cmidrule{2-12}
		&\multirow{3}{*}{\makecell[l]{Few-Shot User\\ Representation Learning 
		}}\!\!\!\!
		&LWA&0.1642 &0.5777 &0.0\% &0.1567 &0.5730 &0.0\% &0.1512 &0.5669 &0.0\% \\
		&&NLBA&0.1632 &0.5664 &-2.0\% &0.1558 &0.5736 &0.8\% &0.1499 &0.5720 &7.6\% \\
		&&JTCN&0.1761 &0.5850 &1.3\% &0.1672 &0.5768 &5.2\% &0.1568 &0.5687 &2.7\% \\
		\cmidrule{2-12}
		&\multirow{3}{*}{\makecell[l] {Optimization-Based \\Meta-Learning}} 
		&MeLU&0.1709 &0.5971 &3.4\% &0.1614 &\underline{0.6219} &\underline{66.9\%} &0.1522 &\underline{0.6307} &\underline{95.3\%} \\
		&&MAMO&\underline{0.1612} &0.5880 &1.8\% &\underline{0.1536} &0.6130 &54.7\% &\underline{0.1488} &0.6210 &80.9\% \\
		&&Meta-Embedding\!\!\!\!&0.1717 &\underline{0.6000} &\underline{3.9\%} &0.1639 &0.6051 &44.0\% &0.1548 &0.6020 &52.4\% \\
		\cmidrule{2-12}
		&
		\multirow{2}{*}{Proposed Methods 
		}&RESUS$_{NN}$&\textbf{0.1601}$^*$ &\underline{0.6086}$^*$ &\underline{5.3\%} &\underline{0.1527}$^*$ &\underline{0.6250} &\underline{71.2\%} &\underline{0.1488} &\textbf{0.6335}$^*$& \textbf{99.5\%} \\
		&&RESUS$_{RR}$&\underline{0.1604}$^*$ &\textbf{0.6109}$^*$ &\textbf{5.8\%} &\textbf{0.1527}$^*$ &\textbf{0.6272}$^*$ &\textbf{74.2\%} &\textbf{0.1470}$^*$ &\underline{0.6276}$^*$ &\underline{90.8\%} \\
		\bottomrule[1pt]
	\end{tabular}%
  }
\end{table*}%

\eat{
\begin{table}[t]
	\centering 
	%	\small
	\caption{Results from online A/B testing (RQ1).}
	\label{tab:online_results}
	%	\resizebox{\linewidth}{!}{
	\begin{tabular}{cccc}
		\toprule[1pt]
		\textbf{Model}&\textbf{CTR (\%)}&\bf \model\%Impr&\bf $p$-value\\
		\midrule[0.5pt] 
		xDeepFM &$4.229\pm 1.898$&29.58\%&.006\\
		%				\midrule[0.5pt] 
		DIN~\cite{DIN}&\underline{$4.720\pm 1.537$}&16.10\%&.036\\
		%				\midrule[0.5pt] 
		DIEN~\cite{DIEN}&\underline{$4.848\pm 1.492$}&13.04\%&.061\\
		\midrule[0.5pt] 
		\makecell{\modelns$_{NN}$\\ (xDeepFM)}&$\mathbf{5.480\pm 1.411}$& / & /\\ 
		%		\midrule[0.5pt] 
		%		\%Improvement& 13.04\%\\
		%		$p$-value&\\
		\bottomrule[1pt]
	\end{tabular} 
	%	}
\end{table}
}

%\begin{figure*}[t]
%	\centering
%	\subcaptionbox{Movielens.\label{fig:hyper:a}}
%	{\includegraphics[width=.32\linewidth]{figures/movielens.pdf}}
%	\subcaptionbox{Frappe.\label{fig:hyper:b}}
%	{\includegraphics[width=.32\linewidth]{figures/frappe}}
%	\subcaptionbox{Taobao.\label{fig:hyper:c}}
%	{\includegraphics[width=.32\linewidth]{figures/taobao}}
%	\caption{Effects of the shared predictor $\Psi$ and pretraining (RQ2).}\label{fig:hyper}
%\end{figure*} 

Table~\ref{tab:main_result} compares different methods on the Logloss and AUC performance over three public datasets.
First, we can see that our proposed \model approaches achieve the best performance on almost all the cases. 
Exceptionally, in the Cold Start-\uppercase\expandafter{\romannumeral1} stage on Frappe, \model lags a little behind the best performing baseline Meta-Embedding on AUC. In fact, Meta-Embedding achieves better AUC performance than MeLU and MAMO in Cold Start-\uppercase\expandafter{\romannumeral1} on all the datasets. 
We {conjecture} the generated user embeddings based on user features acquire useful meta-knowledge that benefits CTR predictions on extremely cold users. 
We can also observe that the performance of our \model approaches is more stable over different datasets than the competitive optimization-based meta-learning methods. The reasons are two-fold. (1) \model performs metric-learning and does not rely on meta-learner to initialize base-learner’s parameters or input embeddings. In contrast, optimization-based meta-learning methods easily suffer from poor initializations produced by the meta-learner. (2)  The matching mechanism in metric-learning distinguishes the utility of each support instance w.r.t. a query instance. Specifically, RESUS can assign lower importance weights to less relevant or noisy support instances, while optimization-based meta-learning approaches treat all the support instances equally.
Second, on average, RESUS$_{NN}$ achieves the best performance on Frappe and Taobao, and it outperforms the most competitive baselines by achieving an average {1.8}\% and {5.0}\% improvements on $RelaImpr$ of AUC, respectively. \modelns$_{RR}$ performs best on Movielens and achieves an average {2.9}\% improvement on $RelaImpr$ of AUC, compared with the most competitive baseline MAMO. 
The results demonstrate that: (1) the decoupling of basis user preferences and residual user preferences contribute to the final prediction performance on cold users;
%the decoupled learning algorithms in \model are effective to capture global user preference knowledge and predict residual preferences using user-specific behavior data; 
(2) both of our proposed base-learners (i.e., NN and RR predictors) are feasible, which obtain similar performance improvements over the existing meta-learning approaches. 
Third, we observe that optimization-based meta-learning methods generally perform better than few-shot user representation learning methods, showing the advantages of performing fine-tuning with user-specific data on new tasks. Note that the Logloss results may not be always consistent with the AUC results because AUC is less sensitive to anomalies and more closely related to model's online ranking performance than Logloss. 
Fourth, among the three optimization-based meta-learning methods, Meta-Embedding generally performs worse than MeLU and MAMO in the Cold Start-\uppercase\expandafter{\romannumeral2}/\uppercase\expandafter{\romannumeral3} stages. 
Meta-Embedding only refines user embeddings (via optimizing the parameters of the embedding generator) based on support sets. On the contrary, MeLU and MAMO finetune all the model parameters based on support sets and hence benefit more from larger support sets.
Finally, regarding the three stages, almost all the methods achieve higher AUC and lower Logloss on larger support sets. This is reasonable because more support samples shall contribute to more accurate user preference estimations. However, few-shot representation learning methods gain little performance improvement in latter stages. The reason is that they apply average pooling, attention mechanism or capsule clustering to encode support instances into a condensed user representation vector, thus risking information loss and yielding suboptimal performance.

\begin{table*}[t]
	\centering
	 \caption{Effects of the shared predictor $\Psi$ and pretraining{, where the bold values are the best results and the underlined values are the most competitive results(RQ2). * indicates a statistically significant level $p$-value<0.05 comparing \model with \model (w/o pretrain).}}
	 \label{tab:hyper1}
 %	\vspace{-0.1in}
	 \resizebox{\textwidth}{!}{
		\begin{tabular}{clccccccccc}
		\toprule
		\multirow{8}{*}{\begin{sideways}{\bf Movielens}\end{sideways}}&  
		\multirow{2}{*}{\bf Methods} & \multicolumn{3}{c}{\bf Cold Start-\uppercase\expandafter{\romannumeral1}} & \multicolumn{3}{c}{\bf Cold Start-\uppercase\expandafter{\romannumeral2}} & \multicolumn{3}{c}{\bf Cold Start-\uppercase\expandafter{\romannumeral3}} \\
					&     &\bf Loss &\bf AUC   &\bf RelaImpr &\bf Loss&\bf AUC   &\bf RelaImpr &\bf Loss&\bf AUC   & \bf RelaImpr \\
		\cmidrule{2-11}
		& MUS &0.4125 &0.5969 &0.0\% &0.4030 &0.6677 &0.0\% &0.4060 &0.6771 &0.0\% \\
		& RESUS$_{NN}$ (w/o pretrain) &0.3372 &0.7580 &166.2\% &0.3323 &0.7714 &61.8\% &0.3344 &0.7735 &54.4\% \\
		& RESUS$_{NN}$ &\underline{0.3322}$^*$ &\underline{0.7638}$^*$ &\underline{172.1\%} &\textbf{0.3274}$^*$ &\underline{0.7745}$^*$ &\underline{63.7\%} &\textbf{0.3306}$^*$ &\underline{0.7771}$^*$ &\underline{56.5\%} \\
		& RESUS$_{RR}$ (w/o pretrain) &0.3385 &0.7568 &164.9\% &0.3317 &0.7702 &61.1\% &0.3325 &0.7747 &55.1\% \\
		& RESUS$_{RR}$ &\textbf{0.3318}$^*$ &\textbf{0.7645}$^*$ &\textbf{172.8\%} &\underline{0.3280}$^*$ &\textbf{0.7772}$^*$ &\textbf{65.3\%} &\underline{0.3314} &\textbf{0.7793}$^*$ &\textbf{57.7\%}\\
		 \midrule
		\multirow{8}{*}{\begin{sideways}{\bf Frappe}\end{sideways}}&  
		\multirow{2}{*}{\bf Methods} & \multicolumn{3}{c}{\bf Cold Start-\uppercase\expandafter{\romannumeral1}} & \multicolumn{3}{c}{\bf Cold Start-\uppercase\expandafter{\romannumeral2}} & \multicolumn{3}{c}{\bf Cold Start-\uppercase\expandafter{\romannumeral3}} \\
		&     &\bf Loss &\bf AUC   &\bf RelaImpr &\bf Loss&\bf AUC   &\bf RelaImpr &\bf Loss&\bf AUC   & \bf RelaImpr \\
		\cmidrule{2-11}
		& MUS & 0.3612 & 0.6477 & 0.0\% & 0.3235 & 0.7604 & 0.0\% & 0.2838 & 0.8531 & 0.0\% \\
		& RESUS$_{NN}$ (w/o pretrain) &0.3305$^*$ &\textbf{0.8687}$^*$ &\textbf{149.6\%} &0.2869 &\underline{0.8886}$^*$ &\underline{49.2\%} &0.2613 &\underline{0.8987}$^*$ &\underline{12.9\%} \\
		& RESUS$_{NN}$ & 0.3322 & 0.8542 & 139.8\% &  \underline{0.2842}$^*$ & 0.8746 & 43.9\% &  \underline{0.2561}$^*$ & 0.8931 & 11.3\%\\
		& RESUS$_{RR}$ (w/o pretrain) &\underline{0.3276} &\underline{0.8548} &\underline{140.2\%} &\textbf{0.2631}$^*$ &\textbf{0.8886}$^*$ &\textbf{49.3\%} &\textbf{0.2406}$^*$ &\textbf{0.9031}$^*$ &\textbf{14.2\%} \\
		& RESUS$_{RR}$ & \textbf{0.3275} & 0.8524 & 138.6\% & 0.3106 & 0.8669 & 40.9\% & 0.2896 & 0.8713 & 5.2\% \\
		 \midrule
		\multirow{8}{*}{\begin{sideways}{\bf Taobao}\end{sideways}}&  
		\multirow{2}{*}{\bf Methods} & \multicolumn{3}{c}{\bf Cold Start-\uppercase\expandafter{\romannumeral1}} & \multicolumn{3}{c}{\bf Cold Start-\uppercase\expandafter{\romannumeral2}} & \multicolumn{3}{c}{\bf Cold Start-\uppercase\expandafter{\romannumeral3}} \\
		&     &\bf Loss &\bf AUC   &\bf RelaImpr &\bf Loss&\bf AUC   &\bf RelaImpr &\bf Loss&\bf AUC   & \bf RelaImpr \\
		\cmidrule{2-11}
		& MUS &0.1622 &0.5761 &0.0\% &0.1550 &0.5969 &0.0\% &0.1494 &0.6041 &0.0\% \\
		& RESUS$_{NN}$ (w/o pretrain) &0.1606 &0.6008 &32.4\% &0.1535 &0.6117 &15.3\% &0.1480 &0.6122 &7.9\% \\
		& RESUS$_{NN}$ &\textbf{0.1601}$^*$ &\underline{0.6086}$^*$ &\underline{42.6\%} &\underline{0.1527} &\underline{0.6250}$^*$ &\underline{29.0\%} &0.1488 &\textbf{0.6335}$^*$ &\textbf{28.3\%} \\
		& RESUS$_{RR}$ (w/o pretrain) &\underline{0.1602} &0.6085 &42.5\% &0.1530 &0.6188 &22.7\% &\underline{0.1474} &0.6190 &14.4\% \\
		& RESUS$_{RR}$ &0.1604 &\textbf{0.6109}$^*$ &\textbf{45.7\%} &\textbf{0.1527} &\textbf{0.6272}$^*$ &\textbf{31.3\%}$^*$&\textbf{0.1470} &\underline{0.6276}$^*$ &\underline{22.7\%} \\
		\bottomrule
		\end{tabular}%
	 }
 %\vspace{-0.1in}
 \end{table*}% 

\subsection{Effects of Key Components (RQ2)}
\label{sec:rq2}

%The advantages of our \model approach lie in two aspects: capturing global preference knowledge and learning residual user preferences. 
%\textcolor{red}{\sout{We now evaluate the key components in RESUS using the following three settings}}
In this section, we conduct experiments to evaluate the key components in \modelns, including (i) the effects of the shared predictor, (ii) the effects of pretraining the shared predictor, and (iii) the effects of the metric-learning counterpart.

First, we remove the shared preference predictor $\Psi$ by always treating its predictions as zero, and perform meta-learning to infer user preferences. The resultant method is exactly our basic metric-based meta-learning approach MUS (in Section~\ref{sec:mus}).  
Table~\ref{tab:hyper1} shows the results of MUS and \model with two different base-learners (i.e., $NN$, $RR$) on three datasets. 
We can see that \model outperforms MUS by a large margin.
%\sout{On average, \modelns$_{NN}$ achieves 84.6\%, 47.4\% and 27.7\%  $RelaImpr$ of AUC for Movielens, Frappe and Taobao, respectively. \modelns$_{RR}$ achieves 85.8\%, 43.3\% and 31.1\% $RelaImpr$ of AUC for Movielens, Frappe and Taobao, respectively.} 
On average, \modelns$_{NN}$ (\modelns$_{RR}$) achieves 84.6\% (85.8\%), 47.4\% (43.3\%) and 32.5\% (32.0\%) $RelaImpr$ of AUC on Movielens, Frappe and Taobao datasets, respectively.
This is because MUS fails to capture the global preference knowledge among different users and the performance suffers from the sparse and insufficient user historical interactions in the support set. 
%from the shared binary labels.
%As described before, MUS fails to capture the global preference knowledge among different users due to the shared binary labels. The results illustrate that the basis user preferences learned by $\Psi$ are important for few-shot CTR prediction tasks. 

Second, we train the shared predictor $\Psi$ together with the other components in \model rather than pretrain it.
% before training \modelns.
%Specifically, we treat $\Psi$ as a parametric module belonging to the meta-learner and optimize its parameters together with those in $\Phi$ and $\Lambda$ through meta-training. 
Specifically, the parameters in $\Psi$ can be viewed as meta-parameters which are optimized with the paremeters in $\Phi$ and $\Lambda$ over meta-train tasks. 
The resultant method is denoted as \model(w/o pretrain). 
Table~\ref{tab:hyper1} shows the effects of two training algorithms applied on $\Psi$. In general, \model achieves better or comparable performance than \model(w/o pretrain). The reason is that the randomly initialized shared predictor $\Psi$ may introduce noises to the target residual user preferences (in Eq.~(\ref{eq:residual})).
%, especially in the early training epochs.
% that obstruct the optimization of the residual predictor $\Lambda$. 
This can sometimes lead to serious problems to the optimization of the residual user preference predictor $\Lambda$. For example, in the stages of Cold Start-\uppercase\expandafter{\romannumeral1}/\uppercase\expandafter{\romannumeral2} on Taobao, performance degrades significantly without pretraining the shared predictor.
It is also interesting to see that in most cases, \model(w/o pretrain) achieves better or comparable performance to the most competitive baselines (referring to Table~\ref{tab:main_result}), and it even outperforms the best performing method, Meta-Embedding, in the Cold Start-\uppercase\expandafter{\romannumeral1} stage on Frappe. These results further confirm the significance of explicitly learning global preference knowledge in addition to residual user preferences.

\begin{table*}[t]
	\centering
	\caption{Effects of different architectures of $\Psi$ and $\Phi$ and gradient boosting on Movielens, where the bold values are the best results and the underlined values are the most competitive results (RQ2\&RQ3). All the improvements of RESUS are statistically significant with $p$-value < 0.01.}
	\label{tab:base_model}
  \resizebox{\textwidth}{!}{
	\begin{tabular}{clccccccccc}
		\toprule
		\multirow{22}[17]{*}{\begin{sideways}\bf Movielens\end{sideways}}&  
		\multirow{2}{*}{\bf Method} & \multicolumn{3}{c}{\bf Cold Start-\uppercase\expandafter{\romannumeral1}} & \multicolumn{3}{c}{\bf Cold Start-\uppercase\expandafter{\romannumeral2}} & \multicolumn{3}{c}{\bf Cold Start-\uppercase\expandafter{\romannumeral3}} \\
		&     & \bf Logloss & \bf AUC   & \bf RelaImpr & \bf Logloss&\bf  AUC   & \bf RelaImpr &\bf Logloss& \bf AUC   & \bf RelaImpr \\
		\cmidrule{2-11}
		& LR &0.3635 &0.7535 &0.0\% &0.3670 &0.7519 &0.0\% &0.3731 &0.7505 &0.0\% \\
		& LR (boosting) &0.3829 &0.7541 &0.2\% &0.3864 &0.7525 &0.2\% &0.3923 &0.7514 &0.3\% \\
		& RESUS$_{NN}$ (LR) &\underline{0.3607} &\underline{0.7615} &\underline{3.2\%} &\underline{0.3599} &\underline{0.7713} &\underline{7.7\%} &\underline{0.3648} &\underline{0.7728} &\underline{8.9\%} \\
		& RESUS$_{RR}$ (LR) &\textbf{0.3584} &\textbf{0.7647} &\textbf{4.4\%} &\textbf{0.3559} &\textbf{0.7773} &\textbf{10.1\%} &\textbf{0.3591} &\textbf{0.7801} &\textbf{11.8\%} \\
		
		\cmidrule{2-11}
		& FM &0.3636 &0.7544 &0.0\% &0.3672 &0.7528 &0.0\% &0.3732 &0.7514 &0.0\% \\
		& FM (boosting) &0.3838 &0.7547 &0.1\% &0.3875 &0.7534 &0.2\% &0.3946 &0.7521 &0.3\% \\
		& RESUS$_{NN}$ (FM) &\underline{0.3605} &\underline{0.7624} &\underline{3.1\%} &\underline{0.3580} &\underline{0.7726} &\underline{7.8\%} &\underline{0.3612} &\underline{0.7746} &\underline{8.9\%} \\
		& RESUS$_{RR}$ (FM) &\textbf{0.3592} &\textbf{0.7647} &\textbf{4.1\%} &\textbf{0.3564} &\textbf{0.7760} &\textbf{9.2\%} &\textbf{0.3607} &\textbf{0.7773} &\textbf{10.0\%} \\
		
		\cmidrule{2-11}
		& Wide\&Deep &0.3641 &0.7525 &0.0\% &0.3677 &0.7508 &0.0\% &0.3738 &0.7494 &0.0\% \\
		& Wide\&Deep (boosting) &0.3817 &0.7550 &1.0\% &0.3880 &0.7533 &1.0\% &0.3906 &0.7526 &1.3\% \\
		& RESUS$_{NN}$ (Wide\&Deep) &\underline{0.3358} &\underline{0.7574} &\underline{1.9\%} &\underline{0.3363} &\underline{0.7654} &\underline{5.8\%} &\underline{0.3420} &\underline{0.7653} &\underline{6.4\%} \\
		& RESUS$_{RR}$ (Wide\&Deep) &\textbf{0.3325} &\textbf{0.7637} &\textbf{4.4\%} &\textbf{0.3275} &\textbf{0.7765} &\textbf{10.2\%} &\textbf{0.3293} &\textbf{0.7799} &\textbf{12.2\%} \\
		
		\cmidrule{2-11}
		& DeepFM &0.3643 &0.7527 &0.0\% &0.3679 &0.7510 &0.0\% &0.3740 &0.7497 &0.0\% \\
		& DeepFM (boosting) &0.3819 &0.7539 &0.5\% &0.3855 &0.7525 &0.6\% &0.3900 &0.7514 &0.7\% \\
		& RESUS$_{NN}$ (DeepFM) &\underline{0.3322} &\underline{0.7638} &\underline{4.4\%} &\underline{0.3274} &\underline{0.7745} &\underline{9.4\%} &\underline{0.3306} &\underline{0.7771} &\underline{11.0\%} \\
		& RESUS$_{RR}$ (DeepFM) &\textbf{0.3318} &\textbf{0.7645} &\textbf{4.7\%} &\textbf{0.3280} &\textbf{0.7772} &\textbf{10.4\%} &\textbf{0.3314} &\textbf{0.7793} &\textbf{11.9\%} \\
		
		\cmidrule{2-11}
		& xDeepFM &0.3633 &0.7550 &0.0\% &0.3669 &0.7533 &0.0\% &0.3728 &0.7519 &0.0\% \\
		& xDeepFM (boosting) &0.3796 &0.7556 &0.3\% &0.3851 &0.7545 &0.4\% &0.3912 &0.7532 &0.5\% \\
		& RESUS$_{NN}$ (xDeepFM) &\underline{0.3312} &\underline{0.7655} &\underline{4.1\%} &\underline{0.3269} &\underline{0.7767} &\underline{9.2\%} &\underline{0.3298} &\underline{0.7790} &\underline{10.7\%} \\
		& RESUS$_{RR}$ (xDeepFM) &\textbf{0.3303} &\textbf{0.7658} &\textbf{4.3\%} &\textbf{0.3262} &\textbf{0.7784} &\textbf{9.9\%} &\textbf{0.3303} &\textbf{0.7798} &\textbf{11.0\%} \\
		
		\cmidrule{2-11}
		% &XGBoost &0.4662 &0.7499 &0.0\% &0.4681 &0.7480 &0.0\% &0.4715 &0.7468 &0.0\% \\
		% &RESUS$_{NN}$ (XGBoost) &\underline{0.3628} &\underline{0.7589} &\underline{3.6\%} &\textbf{0.3588} &\underline{0.7709} &\underline{9.3\%} &\underline{0.3625} &\underline{0.7722} &\underline{10.3\%} \\
		% &RESUS$_{RR}$ (XGBoost) &\textbf{0.3606} &\textbf{0.7625} &\textbf{5.1\%} &\underline{0.3597} &\textbf{0.7727} &\textbf{10.0\%} &\textbf{0.3621} &\textbf{0.7757} &\textbf{11.7\%}\\
		&LightGBM &0.3628 &0.7569 &0.0\% &0.3665 &0.7552 &0.0\% &0.3726 &0.7538 &0.0\% \\
		&RESUS$_{NN}$ (LightGBM) &\underline{ 0.3606} &\underline{ 0.7641} &\underline{ 2.8\%} &\underline{ 0.3623} &\underline{ 0.7708} &\underline{ 6.1\%} &\underline{ 0.3691} &\underline{ 0.7725} &\underline{ 7.4\%} \\
		&RESUS$_{RR}$ (LightGBM) &\textbf{0.3580} &\textbf{0.7675} &\textbf{4.1\%} &\textbf{0.3555} &\textbf{0.7780} &\textbf{8.9\%} &\textbf{0.3590} &\textbf{0.7799} &\textbf{10.3\%}\\
		\bottomrule[1pt]
	\end{tabular}%
	\label{tab:results}%
  }
\end{table*}%

Third, we remove the residual prediction counterpart and debase \model to the shared predictor $\Psi$.
We use $6$ different architectures to implement $\Psi$, i.e., logistic regression (LR), factorization machine (FM), 
Wide\&Deep~\cite{wide_and_deep}, DeepFM~\cite{DeepFM}, xDeepFM~\cite{xDeepFM} {and LightGBM~\cite{LightGBM}}. Since the base predictor is trained with training samples from different users, it is able to learn global preference knowledge including global item characteristics that are discriminative to the prediction results.
%For fair comparison, we report the performance of the complete \model using the same architecture for the shared predictor $\Psi$ and the encoder $\Phi$ accordingly. 
{We report the performance of \model using the same architecture for the shared predictor $\Psi$ and the encoder $\Phi$, except for \model(LightGBM) that employs DeepFM as the encoder.  
%The only exception is that RESUS using LightGBM as $\Psi$ employs encoding layers of DeepFM as $\Phi$.
}
Table~\ref{tab:base_model} provides the results on Movielens, while the same conclusions can be drawn on the other datasets. 
We have three important observations. 
(1) \model improves the prediction performance significantly by augmenting global preferences with user-specific residual preferences. 
{The AUC performance of \model increases with larger support sets. The shared predictor performs the worst on all the cases, which confirms the limitation of supervised learning on cold users. 
%	Note that all the implementations of $\Psi$ show worse performance in latter stages as the query sets are different. In contrast, RESUS benefit from user-specific information and consistently achieves higher AUC an $RelaImpr$ in latter stages.
}
%the conventional CTR prediction methods adopt supervised learning and suffer from the user cold-start challenge. 
(2) Different architectures of $\Psi$ and $\Phi$ effect the performance of \modelns. 
In general, a better architecture of $\Psi$ benefits the final performance of \modelns. For example, on average, xDeepFM outperforms LR by {0.5\%} $RelaImpr$ of AUC, and \modelns$_{RR}$ (xDeepFM) outperforms \modelns$_{RR}$ (LR) by {0.2\%} $RelaImpr$ of AUC. 
Among all the compared architectures, {LightGBM} achieves the best performance. As a result, \modelns$_{NN}$ and \modelns$_{RR}$ based on {LightGBM} also achieve the best Logloss and AUC results. These results verify the generality of \modelns, encouraging more advanced CTR prediction models to be incorporated into \modelns.
(3) The relative performance improvements achieved by \model are quite stable over different shared predictors. This is because the metric-learning counterpart in \model utilizes the shared predictor to derive representations of feature vectors in support and query instances and it also benefits from stronger shared predictors.

\subsection{Comparison with Gradient Boosting (RQ3)}\label{sec:compare_boosting}

We compare \model with the gradient boosting technique over different architectures of $\Psi$. For gradient boosting, we first train a shared predictor $\Psi$ with the training set (i.e., $D_{train}$), and then train a subsequent predictor with the same structure as $\Psi$ to fit the residual errors on the support sets of all the test users. After training, the outputs of the two predictors on the query instances for the test users are added to make the final predictions. Table~\ref{tab:base_model} provides the results on Movielens and the same conclusions can be drawn on the other datasets. From the results, we can see that gradient boosting can improve the AUC of the shared predictor on all the cases. On average, gradient boosting achieves {0.3\%, 0.2\%, 1.1\%, 0.6\% and 0.4\%} $RelaImpr$ of AUC for LR, FM, Wide\&Deep, DeepFM and xDeepFM, respectively. We can also see that \model outperforms gradient boosting significantly over all the architectures of $\Psi$. This indicates that the performance gain of \model comes from capturing global preference knowledge and user-specific residual preferences separately instead of model ensemble.
Besides, we notice that gradient boosting incurs higher Logloss than the shared predictor while improving AUC performance. This is because boosting methods aim at improving classification accuracy and may not necessarily lead to lower Logloss results.

\subsection{Robustness Study (RQ4)\label{sec:noise_study}}

\begin{figure*}[htp!]
	\centering
	\subcaptionbox{Logloss, Movielens\label{fig:noise:a}}
	{\includegraphics[width=.48\linewidth]{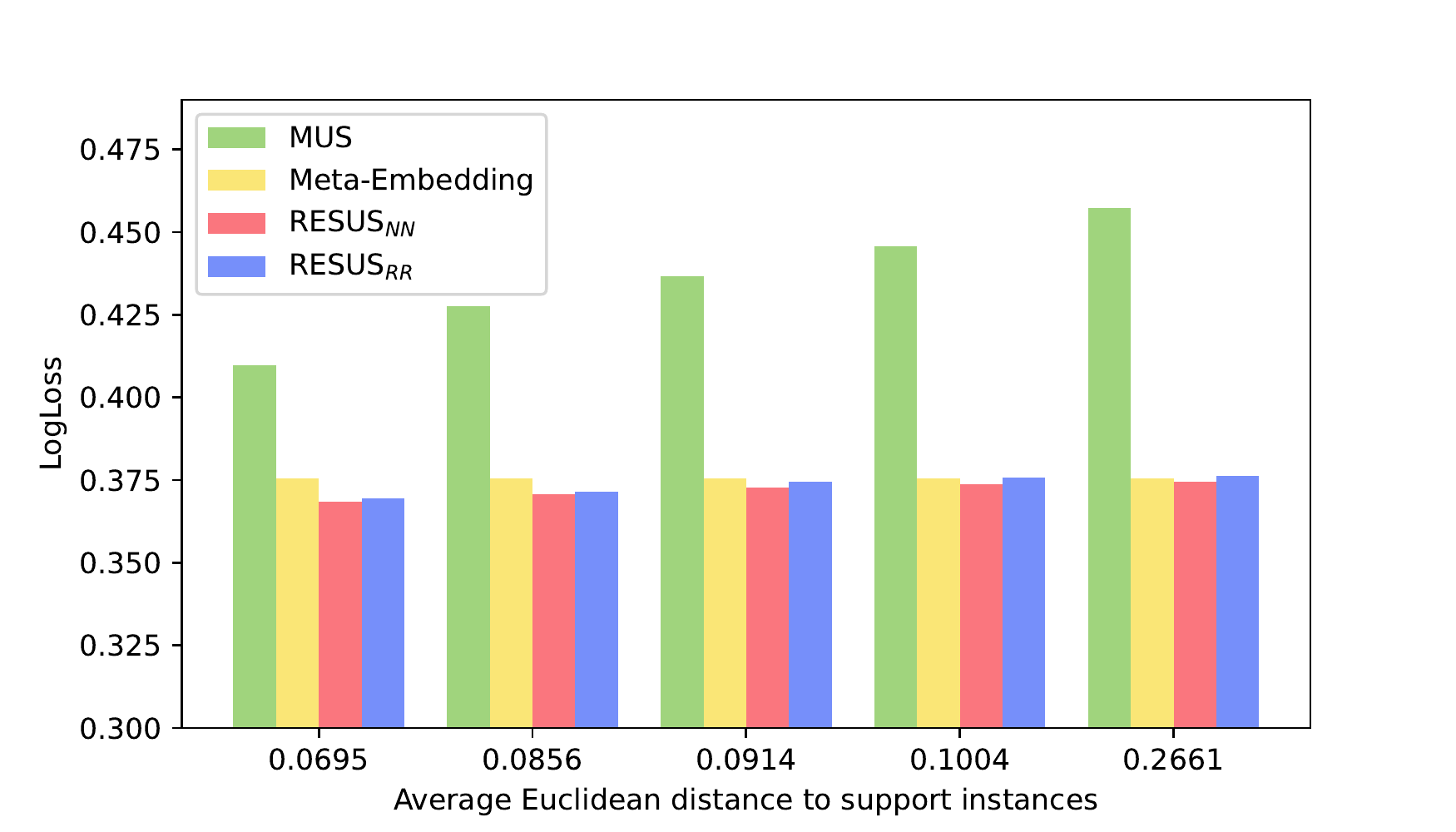}}
	\subcaptionbox{AUC, Movielens\label{fig:noise:b}}
	{\includegraphics[width=.48\linewidth]{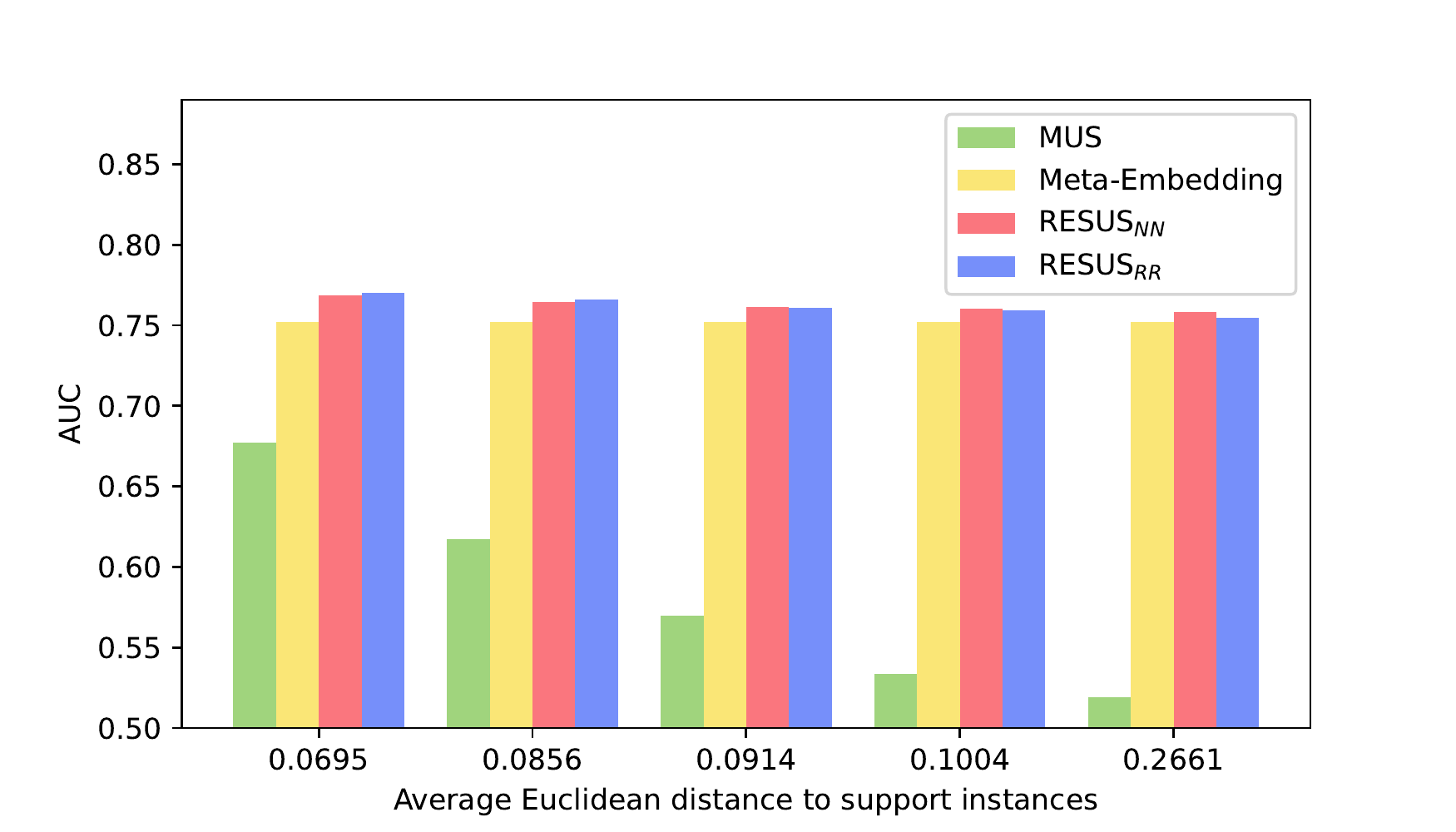}}
%	\subcaptionbox{Logloss, Frappe\label{fig:noise:c}}
%	{\includegraphics[width=.48\linewidth]{figures/noise_loss_frappe.png}}
%	\subcaptionbox{AUC, Frappe\label{fig:noise:d}}
%	{\includegraphics[width=.48\linewidth]{figures/noise_auc_frappe.png}}
	\subcaptionbox{Logloss, Taobao\label{fig:noise:c}}
	{\includegraphics[width=.48\linewidth]{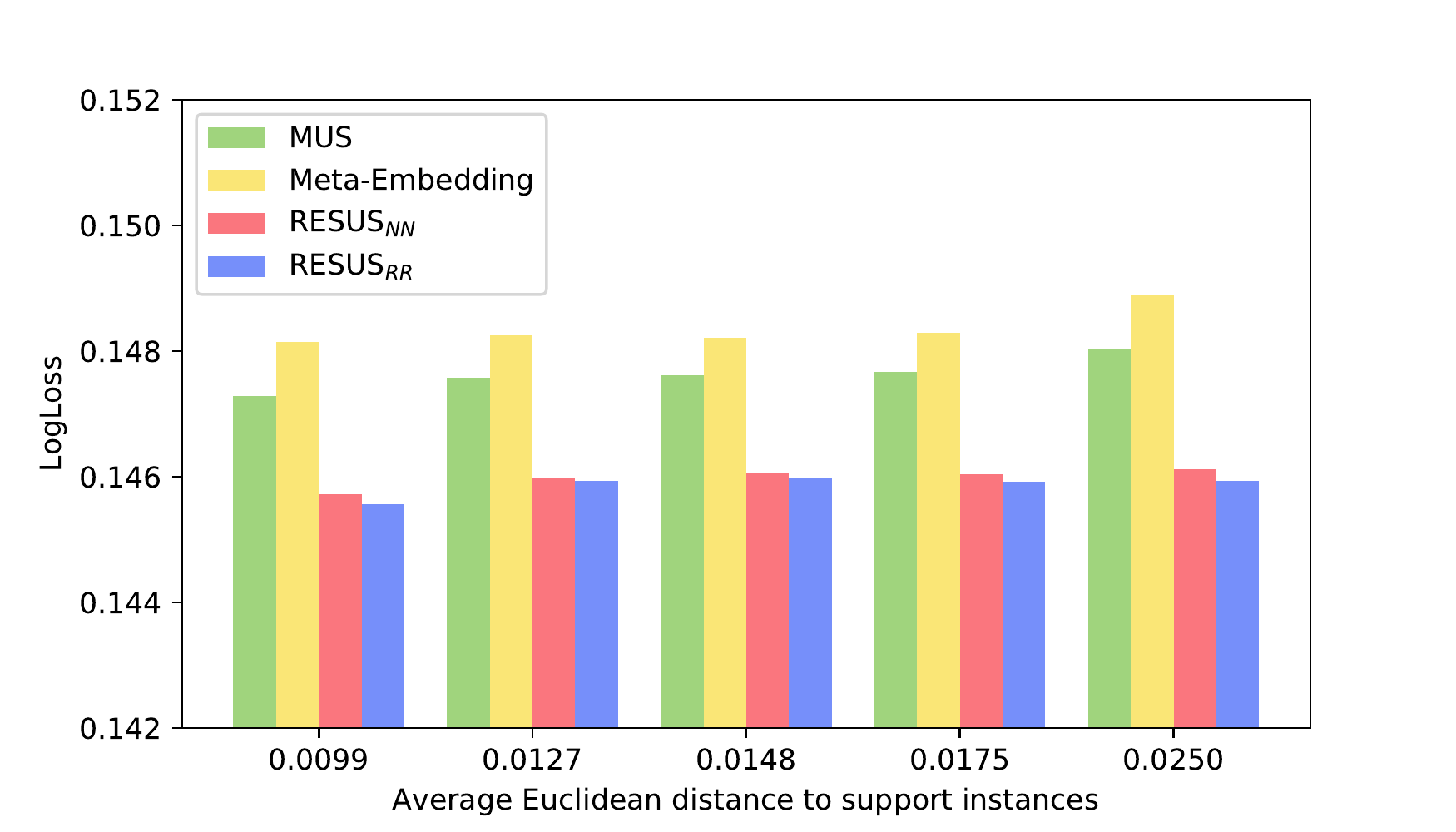}}
	\subcaptionbox{AUC, Taobao\label{fig:noise:d}}
	{\includegraphics[width=.48\linewidth]{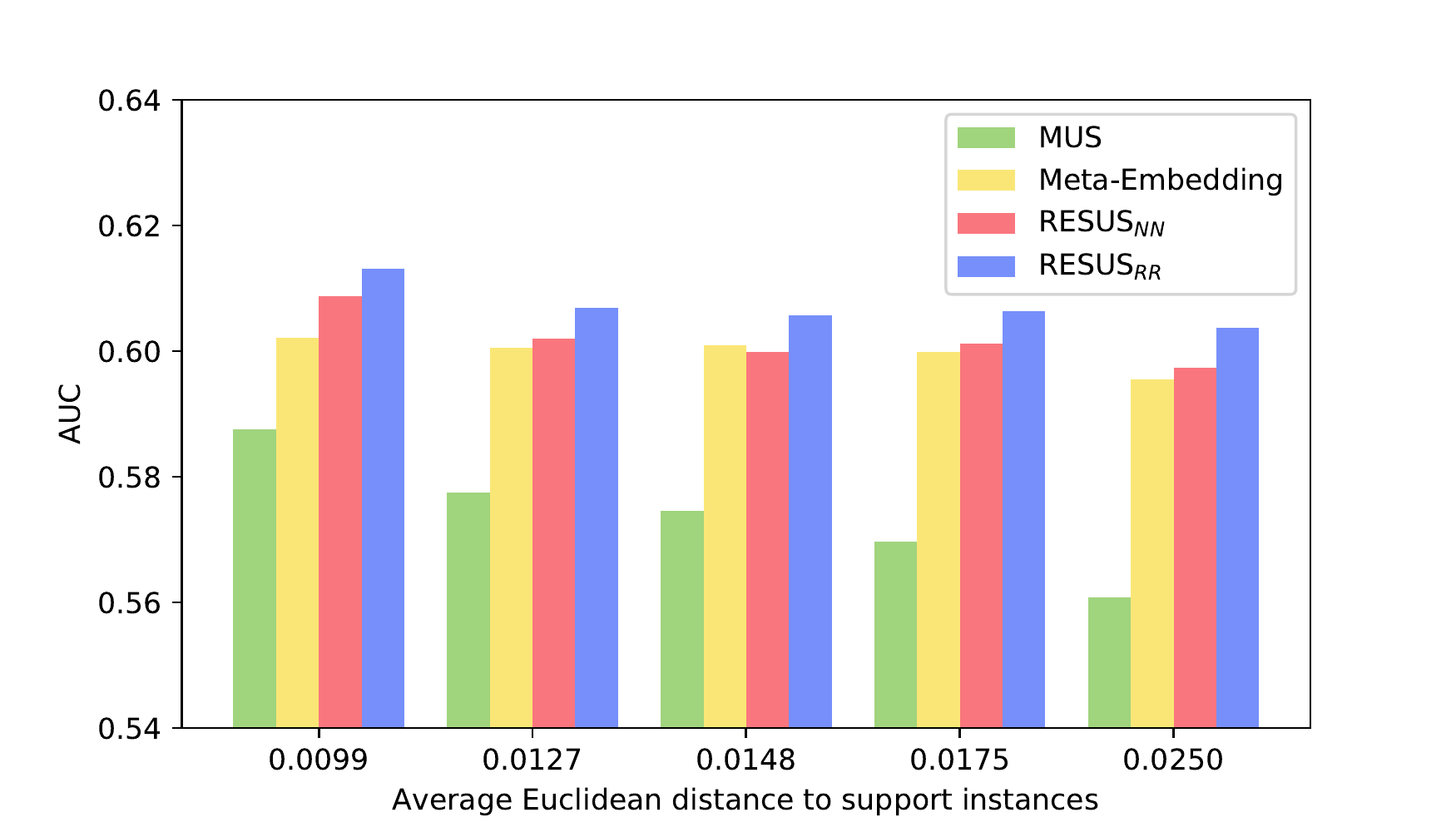}}
	\caption{{Robustness study results on different support sets (RQ4).}}\label{fig:noise}
\end{figure*} 

%\begin{figure*}[htp!]
%	\centering
%	\subcaptionbox{LogLoss, Movielens\label{fig:noise2:a}}
%	{\includegraphics[width=.48\linewidth]{figures/noise_loss_frappe2.png}}
%	\subcaptionbox{AUC, Movielens\label{fig:noise2:b}}
%	{\includegraphics[width=.48\linewidth]{figures/noise_auc_frappe2.png}}
%	\caption{\textcolor{red}{Robustness study on different nearest support instances with five farest support instances (RQ4).}}\label{fig:noise}
%\end{figure*} 
{
As support sets may involve noises or unrelated historical interactions to the predictions of query instances, we now evaluate the robustness of \model on support sets with different noise levels. 
%In this experiment, we focus on the challenging cases in Cold Start-\uppercase\expandafter{\romannumeral1} which correspond to extremely cold users with the smallest support sets. We 
In this experiment, we prepare $5$ meta-test sets with different support sets but the same query sets. 
%Specifically, for each query instance $x_u^Q$ in the meta-test tasks in Cold Start-\uppercase\expandafter{\romannumeral1}, we divide its support set $S_u$ into five equal-sized groups with different noise levels. 
Specifically, for each test user $u$, we sort the historical interactions in time order and consider the first $\tau$ interactions as the \emph{temporary support set}.  
%$S_u$\textcolor{red}{, where $\tau$ is $30$ and $150$ for Movielens and Taobao, respectively}. 
The remaining interactions are formed into the query set $Q_u$.  For each query instance $x_u^Q\in Q_u$, we divide $S_u$ into $5$ equal-sized groups with different noise levels. 
Since we cannot recognize noisy support instances with respect to the query instance $x_u^Q$, we compute Euclidean distance between $x_u^Q$ with every support instance in $S_u$ based on their encoded vector representations. We use the pretrained shared predictor (i.e., DeepFM excluding the last prediction layer) to encode instances which is independent of the meta-learning approaches. We then sort the support instances in {$S_u$} in ascending order based on the distances and divide them into $5$ equal-sized groups, i.e., {$S_u=S_{u,1}\cup\cdots\cup S_{u,5}$}. Intuitively, larger distances imply the support instances are less relevant to the query instance, which would compromise the effectiveness of the matching mechanism in metric-learning. Hence, we regard the groups with larger distances (e.g., $S_{u,5}$) as support sets with higher noise levels to the query instance. By aggregating the groups from all the query instances by the noise level, we obtain $5$ meta-test sets that share the same query sets but the corresponding support sets are disjoint and in different noise levels. Note that all the derived support sets have the same size, i.e., ${\tau}/{5}$ support instances,  corresponding to the CTR prediction tasks on extremely cold users.}

{
Figure~\ref{fig:noise} shows the Logloss and AUC performance of MUS, \model and Meta-Embedding on different meta-test sets from Movielens and Taobao. We exclude the results on Frappe dataset which does not ensure the time order between support sets and query sets. The x-axis reports the average Euclidean distance from query instances to support instances for each of the meta-test sets. According to Table~\ref{tab:main_result}, Meta-Embedding is the best performing baseline in Cold Start-\uppercase\expandafter{\romannumeral1}. We can observe that \model consistently outperforms Meta-Embedding and its performance is stable over the meta-test sets with different noise levels on two datasets. The AUC performance of \model slightly decreases with larger distances because the matching mechanism in metric-learning can be affected by the similarity between support and query instances. Nevertheless, compared with MUS, \model is much more robust to support sets with little utility (i.e., meta-test sets with larger distances). 
%Interestingly, both Meta-Embedding and \model can still achieve good performance when support sets only involves $6$ historical interactions, compared with their performance on the cases in Cold Start-\uppercase\expandafter{\romannumeral1}.
%Interestingly, both Meta-Embedding and \model can still achieve good performance when support sets only involve 1/5 of the support instances in the original cases in the \textcolor{red}{\sout{Cold Start-\uppercase\expandafter{\romannumeral1}} Cold Start-\uppercase\expandafter{\romannumeral3}} stage.
}

\subsection{Time Cost Study (RQ{5})}\label{sec:time_cost}

\begin{table*}[b]  
	%\small
	\centering
	%	\small
	\caption{Empirical time cost (in second) comparison on Movielens, {where the pretraining time cost (in second) is provided in brackets (RQ{5}).}}
	
	% \label{tab:time}
	% \resizebox{\textwidth}{!}{
	% 	\begin{tabular}{llccc}
	% 		\toprule[1pt]
	% 		\bf{Method Class} & \bf{Method} & \multicolumn{1}{l}{\#\bf{Training epochs}} & \multicolumn{1}{l}{\bf{Training Time (s)}} & \multicolumn{1}{l}{\bf{Test Time (s)}} \\
	% 		\midrule[0.5pt]
	% 		\multirow{3}{*}{\makecell[l]{Few-Shot User\\ Representation Learning 
	% 		}}\!\!\!\! 
	% 		& LWA & 6 & 784 & 19 \\
	% 		& NLBA & 3 & 350 & 20 \\
	% 		& JTCN & 6 & 571 & 16 \\
	% 		\midrule[0.5pt]
	% 		\multirow{3}{*}{\makecell[l]{Optimization-Based \\Meta-Learning}}  
	% 		& MeLU & 2 & 412 & 34 \\
	% 		& MAMO & 10 & 4641 & 104 \\
	% 		& Meta-Embedding & 10 & 1617 (+24) & 22 \\
	% 		\midrule[0.5pt]
	% 		\multirow{4}{*}{\makecell[l]{Proposed Methods} }
	% 		& RESUS$_{NN}$ (w/o pretrain) & 5 & 542 & 10 \\
	% 		& RESUS$_{RR}$ (w/o pretrain) & 3 & 329 & 10\\
	% 		& RESUS$_{NN}$ & 4 & 420 (+20) & 10 \\
	% 		& RESUS$_{RR}$ & \textbf{2} & \textbf{218 (+20)} & \textbf{10}\\
	% 		\bottomrule[1pt]
	% \end{tabular}}
	
	\label{tab:time}
	\resizebox{\textwidth}{!}{
		\begin{tabular}{llccc}
			\toprule[1pt]
			\bf{Method Class} & \bf{Method} & \multicolumn{1}{l}{\#\bf{Training epochs}} & \multicolumn{1}{l}{\bf{Training Time (s)}} & \multicolumn{1}{l}{\bf{Test Time (s)}} \\
			\midrule[0.5pt]
			\multirow{3}{*}{\makecell[l]{Few-Shot User\\ Representation Learning 
			}}\!\!\!\! 
			& LWA & 8 & 1085 & 19 \\
			& NLBA & 5 & 661 & 20 \\
			& JTCN & 8 & 761 & 16 \\
			\midrule[0.5pt]
			\multirow{3}{*}{\makecell[l]{Optimization-Based \\Meta-Learning}}  
			& MeLU & 4 & 820 & 34 \\
			& MAMO & 10 & 4641 & 104 \\
			& Meta-Embedding & 10 & 1617 (+24) & 22 \\
			\midrule[0.5pt]
			\multirow{4}{*}{\makecell[l]{Proposed Methods} }
			& RESUS$_{NN}$ (w/o pretrain) & 7 & 758 & 10 \\
			& RESUS$_{RR}$ (w/o pretrain) & 5 & 547 & 10\\
			& RESUS$_{NN}$ & 6 & 628 (+20) & 10 \\
			& RESUS$_{RR}$ & \textbf{4} & \textbf{438 (+20)} & \textbf{10}\\
			\bottomrule[1pt]
	\end{tabular}}
	
	%	\begin{tabular}{l|c|c|c|c}
	%		\toprule[1pt]
	%		\textbf{Method}
	%		&\textbf{t$_{train}$/Epoch}&\textbf{\#Epoch}&\textbf{t$_{train}$}&\textbf{t$_{test}$}\\
	%		\midrule[0.5pt]
	%		MAMO& 452&10&4520&50\\ 
	%		MeLU& 195&10&1950&23\\ 
	%		Meta-Embedding&102&10&1020&11\\ 
	%		JTCN&110&29&3190&10\\ 
	%		NLBA&221&6&1326&4\\ 
	%		LWA&214&5&1070&4\\ 		
	%		RESUS$_{NN}$&201&4&801&4\\ 
	%		RESUS$_{RR}$&210&3&630&4\\ 
	%		\bottomrule[1pt]
	%	\end{tabular}  
\end{table*} 

% \begin{figure}
% 	\label{fig:noise}
% 	\title{The capacity of resistence to noise in support set (RQ5).}
% 	\includegraphics[width=\linewidth]{figures/mae.png}
% \end{figure}

\begin{figure}[t]
	\centering
	\includegraphics[width=0.7\linewidth]{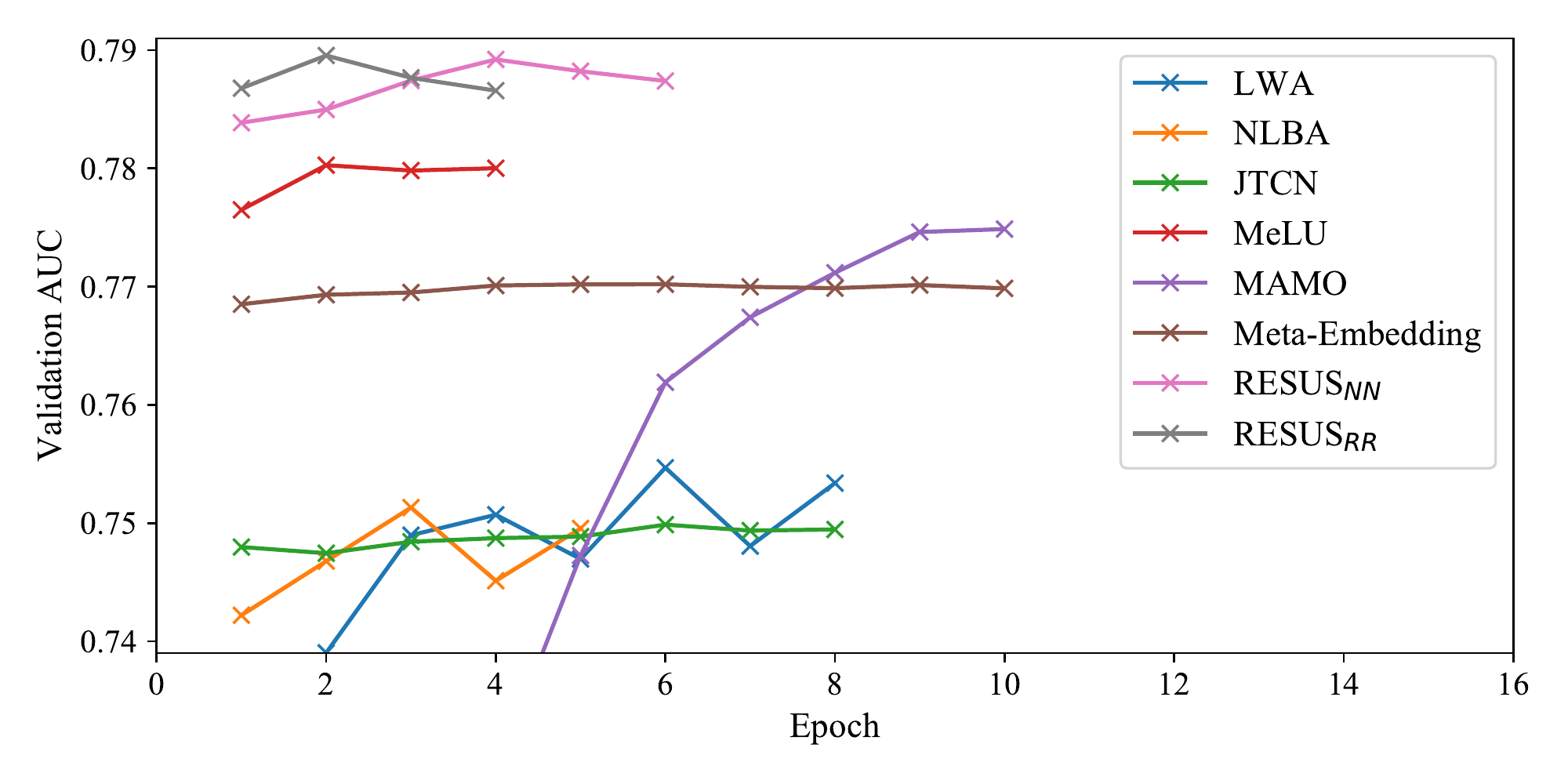}
	\caption{{Converge speed of different models on Movielens (RQ{5}).}}
	\label{fig:epoch}
\end{figure}

We evaluate the empirical time cost of different methods. Table~\ref{tab:time} reports the training time t$_{train}$ and testing time t$_{test}$ using Movielens dataset. The relative performance on training and test time is the same on Frappe and Taobao, and we eliminate the results to avoid redundancy. 
In Table~\ref{tab:time}, it is easy to see that our \model approaches achieve the highest training efficiency in terms of t$_{train}$. The main reason is that the NN/RR predictor in \model does not require gradient-based fine-tuning on every task. 
The training time of optimization-based meta-learning methods is generally higher than that of few-shot user preference learning methods. 
Besides, \model approaches report the lowest testing time, which can perform inference as fast as few-shot user representation learning methods. This is a desirable property in real large-scale recommender systems. 
We also observe that the training and testing efficiency of the same method may not be consistent. For example, {Meta-Embedding} reports higher t$_{train}$ than MeLU due to more training epochs, but takes lower t$_{test}$ than MeLU. In practice, the performance on t$_{test}$ is much more crucial since it directly affects online serving latency and system throughput. 
% We exclude the pretraining of $\Psi$ and focus on the meta-learning process.
%Note that we perform early-stopping during training and hence different methods take different numbers of training epochs. 
{Figure~\ref*{fig:epoch} shows the AUC results on the validation set over training epochs. We perform early stopping when the number of epochs reaches $10$ or the AUC validation performance is getting worse for two consecutive epochs. We can see that LWA, NLBA, JTCN, MeLU, \modelns$_{NN}$ and \modelns$_{RR}$ converge quickly, but MAMO and Meta-Embedding take more training epochs till convergence. The reason is that MAMO starts with randomly initialized memories which needs more updates. Meta-Embedding simultaneously minimizes the prediction error and the mimic loss for fast adaptation, which requires more adjustments for optimal balance between two objectives.}

%}

%We report results on Movielens dataset, including training time cost t$_{train}$ and testing time cost t$_{test}$. From the results, we can see that RESUS methods achieves the lowest time cost in both training and testing. Specifically, the time costs of RESUS methods are similar to few-shot user representation learning methods LWA and NLBA, and are much lower than optimization-based meta-learning methods MAMO, MeLU and Meta-Embedding. This is because the computation flow of RESUS base learner is fully feed-forward, without gradient-based optimization loops in optimization-based meta-learning methods. Thus RESUS can achieve similar inference time efficiency to few-shot user representation learning methods, and is suitable for large-scale real-world recommendation scenarios. Besides, we observe that the comparison of the training costs of different methods does not strictly follow the comparison of testing costs. For example, JTCN has lower testing time cost than MeLU, while has higher training time cost than MeLU due to more training epochs. But in real applications, the comparison on testing time costs is more important since it directly affects online serving latency and system throughputs. 

\section{Related Work}\label{sec:relate}

%\subsection{Cold-Start Recommendation} 
Our paper is related with works on cold-start recommendation. 
While there can be user cold-start and item cold-start problems, most of the proposed methods can be applied to either problem. We thus describe the works from the perspective of user cold-start recommendation. 
Existing literature focused on two types of cold users: completely-cold users that have no interaction data available, and cold users that have very limited historical behavior data. For the first type, many researches~\cite{DBLP:conf/recsys/BarkanKYK19,DBLP:conf/aaai/XuZCLS20,DBLP:conf/sigir/Hansen0SAL20,DBLP:conf/aaai/LiJL00H19} relied on the side information to learn the representation of a target user. %The networks are typically trained with behavior data from non-cold users. 
For the second type, some studies considered the collaborative filtering (CF) setting where only the user-item interaction or rating matrix is available, and proposed data imputation techniques to fill in empty entries in the extremely sparse rows/columns. 
For example, zero-injection~\cite{DBLP:conf/icde/HwangPKLL16, DBLP:journals/tkde/LeeHPLKL19} was proposed to find uninteresting items and then inject zero ratings to them as negative samples. Recent works~\cite{DBLP:conf/www/ChaeKKC19, DBLP:conf/sigir/ChaeKCK20} employed generative adversarial networks to generate plausible ratings to impute the sparse matrix. %These works focus on the CF setting, thus is different from our problem on CTR prediction where user, item and contextual features are accessible and need to be exploited.
Different from the above works, our work aims to improve CTR prediction performance on cold users with very few historical behaviors. To address the problem, some works exploited external user information such as social networks~\cite{DBLP:conf/aaai/0004JCGCA19,DBLP:conf/sigir/LiuOMM20} and cross-domain user behaviors~\cite{DBLP:conf/recsys/AggarwalYK19, DBLP:conf/aaai/FuPWXL19, DBLP:conf/sigir/NazariCPLCVC20, DBLP:conf/sigir/BiSYWWX20a, DBLP:conf/sigir/ZhaoLXDS20, DBLP:journal/tois/HuCCGXY16, DBLP:journal/tois/CaoHNWHMC17}. For example, Hu \emph{et al.}~\cite{DBLP:conf/aaai/0004JCGCA19} proposed to build graph neural networks based on users' social relations and enhance cold user representations by propagating information from neighboring users. Nazari \emph{et al.}~\cite{DBLP:conf/sigir/BiSYWWX20a} proposed to recommend podcasts for cold users by using their music consumption behaviors. These methods, however, rely on specific types of external user knowledge, which may not always be available in the context of CTR prediction. 

When no external user information is accessible, recent works formulated the user cold-start problem as few-shot learning tasks and developed meta-learning approaches, which can be generally categorized into two groups. 
The first group is referred to as \emph{few-shot user representation learning} methods~\cite{DBLP:conf/nips/VartakTMBL17,DBLP:conf/nips/VolkovsYP17,DBLP:conf/www/LiWW020,DBLP:conf/sigir/LiangXYY20}. The key idea is to absorb the limited historical behaviors of the target cold user and generate a fixed-length vector as the user representation, which can be achieved by a parameterized function. The parameters are optimized via episode-based training in order to learn transferrable knowledge across different tasks (i.e,. cold users). Due to the varying number of historical interactions among different users, these methods generate user representations by applying average~\cite{DBLP:conf/nips/VartakTMBL17,DBLP:conf/nips/VolkovsYP17}, attention-based~\cite{DBLP:conf/www/LiWW020}, or capsule clustering-based~\cite{DBLP:conf/sigir/LiangXYY20} pooling methods on the encoded historical interactions of each user, thus risking information loss and yielding suboptimal performance~\cite{DBLP:conf/ijcai/ZhaoWZ17,DBLP:conf/iccv/PassalisT17,gholamalinezhad2020pooling} .
The second group includes optimization-based meta-learning approaches~\cite{DBLP:conf/sigir/PanLATH19, DBLP:conf/ijcnn/Bharadhwaj19,DBLP:conf/kdd/LeeIJCC19,DBLP:conf/kdd/DongYYXZ20,DBLP:conf/kdd/Lu0S20, zhu2021learning}. 
They typically followed the idea of MAML~\cite{DBLP:conf/icml/FinnAL17}, and aimed to learn an initialization scheme for the parameters of the prediction model. %with the objective of fast adapting to new cold users.  
For example, Pan \emph{et al.}~\cite{DBLP:conf/sigir/PanLATH19} proposed to learn an ID embedding generator for cold users, which outputs initial user embeddings based on user-specific features. The initialized user embedding is expected to converge to a good point using very few user's interaction data. Lee \emph{et al.}~\cite{DBLP:conf/kdd/LeeIJCC19} proposed to learn the initialization scheme of the parameters of neural networks, where each network performs predictions for a particular cold user. % so that the network layers can be optimized with the limited training samples of a cold user.
The main drawback of optimization-based meta-learning approaches is the high computation and memory cost caused by the calculation of high-order derivatives in bi-level optimization process and the iterative fine-tuning for every new task~\cite{huisman2020survey,hospedales2020meta}.
In contrast, our \model approach employs two efficient base-learners: the nearest-neighbor predictor and the ridge-regression predictor with closed form solutions, which avoids expensive fine-tuning during both training and test time. Besides, \model employs conventional CTR prediction methods such as DeepFM~\cite{DeepFM} to explicitly capture global preference knowledge, and our research is thus orthogonal to such supervised-learning methods. Furthermore, we have empirically demonstrated the effectiveness and efficiency of \model compared with various existing meta-learning approaches.

\section{Conclusions and Future Work}\label{sec:conclude}

This paper has introduced a generic decoupled preference learning framework named RESUS for CTR prediction on cold users. There are two key insights in this work. 
%First, we propose to decouple the learning of global user preference knowledge from the learning of residual preferences for individual users. The former is achieved by a shared predictor that learns from interactions of different users. The latter is performed by meta-learning that can learn quickly from a few user-specific historical interactions.
First, we propose to decouple user preferences into two parts, namely basis user preferences and residual user preferences. 
The former is inferred by a shared predictor based on input feature vectors of CTR instances. We train the shared predictor in a fully supervised manner such that it acquires global preference knowledge from collective users. 
%is learned by a shared predictor that acquires global preference knowledge from interactions of collective users. 
The latter is inferred based on a few user-specific historical interactions through the matching mechanism in metric-learning. 
The two parts actually complement each other and \model can work well when input features are informative to the prediction or user's historical interactions are useful to transfer label information to the prediction through matching.
%Second, to address the limitations of the existing meta-learning algorithms to few-shot CTR prediction, we devise two kinds of base learners which are efficient and easy to optimize.
Second, we customize meta-learning to our few-shot CTR prediction problem and devise two kinds of base-learners which are efficient and easy to optimize. Extensive experiments on three public datasets have demonstrated that our proposed RESUS approach is generic and efficient, achieving the state-of-the-art CTR prediction performance on cold users. 
To these ends, we believe metric-learning is a promising research direction for cold-start recommendation. 
% There are many interesting problems to be explored for our \modelns. First, we will visualize and analyse the effect of the rescaling coefficient for fusing the two preferences across different cold-start stages. As the belif and bias of two preferences are related to diverse information of specific support sets and query sets, \model can be further improved by some adaptive mechanisms such as gated attention. Second, our \model only focuses on cold users, while the item side is equally significant. We will try to design metric-based approaches for the item cold-start problem. Furthermore, how to perform metric-based meta-learning simultaneously on the two issues is a practical but unexplored problem.

Our work can be extended in multiple directions. In this paper, we fuse the basis and residual user preferences via a simple weighted sum. It is interesting to see how different fusing mechanisms (e.g., the gating mechanism) would affect the final prediction accuracy. Typically, in commercial recommender systems, new interaction data is collected continuously, and a cold user would become warmer gradually. How to adapt RESUS to the change in user coldness is another interesting direction. 

%general meta-learning framework for CTR prediction on cold users. We show that by decoupling user residual preferences from shared knowledge, we can employ a simple yet effective meta-learning approach to make predictions for cold users.The key idea is to adopt a decoupled predictor-encoder framework, where the predictor tries to learn the shared knowledge among different users while the encoder is combined with a base learner to model each user's distinct preference. We propose two choices of base learners, including nearest neighbor and ridge regression, and demonstrate the effectiveness of both choices in the proposed method. We conduct experiments on two public datasets and a real-world recommender system. The experimental results verify the superior performance of our method compared with state-of-the-arts.

\section*{Acknowledgements}
%This work is supported by the Tencent Wechat Rhino-Bird Focused Research Program. Yanyan Shen is also supported by Shanghai Municipal Science and Technology Major Project (2021SHZDZX0102) and the SJTU Global Strategic Partnership Fund (2021 SJTU-HKUST).
The authors would like to thank the anonymous reviewers for their insightful reviews. This  work is supported by the National Key Research and Development Program of China  (2022YFE0200500), Shanghai Municipal Science and Technology Major Project  (2021SHZDZX0102), the Tencent Wechat Rhino-Bird Focused Research Program, and SJTU Global Strategic Partnership Fund (2021 SJTU-HKUST).

\bibliographystyle{ACM-Reference-Format}
\bibliography{kdd-short.bib}

\end{document}